\documentclass[aps,prx,twocolumn,superscriptaddress]{revtex4-2}
\usepackage{amsmath}
\usepackage{graphicx}
\usepackage{dcolumn}
\usepackage{bm}
\usepackage{braket}

\begin{document}


\title{TLS Dynamics in a Superconducting Qubit Due to Background Ionizing Radiation }

\author{Ted Thorbeck}
\email{ted.thorbeck@ibm.com}
\affiliation{IBM Quantum, IBM T.J. Watson Research Center, Yorktown Heights, NY 10598, USA}
\author{Andrew Eddins}
\affiliation{IBM Quantum, MIT-IBM Watson AI lab, Cambridge MA, 02142, USA}
\author{Isaac Lauer}
\affiliation{IBM Quantum, IBM T.J. Watson Research Center, Yorktown Heights, NY 10598, USA}
\author{Douglas T. McClure}
\affiliation{IBM Quantum, IBM T.J. Watson Research Center, Yorktown Heights, NY 10598, USA}
\author{Malcolm Carroll}
\affiliation{IBM Quantum, IBM T.J. Watson Research Center, Yorktown Heights, NY 10598, USA}

\date{\today}

\begin{abstract}

Superconducting qubit lifetimes must be both long and stable to provide an adequate foundation for quantum computing.
This stability is imperiled by two-level systems (TLSs), currently a dominant loss mechanism, which exhibit slow spectral dynamics that destabilize qubit lifetimes on hour timescales.
Stability is also threatened at millisecond timescales, where ionizing radiation has recently been found to cause bursts of correlated multi-qubit decays, complicating quantum error correction.
Here we study both ionizing radiation and TLS dynamics on a 27-qubit processor, repurposing the standard transmon qubits as sensors of both radiation impacts and TLS dynamics.
Unlike prior literature, we observe resilience of the qubit lifetimes to the transient quasiparticles generated by the impact of radiation.
However, we also observe a new interaction between these two processes, “TLS scrambling,” in which a radiation impact causes multiple TLSs to jump in frequency, which we suggest is due to the same charge rearrangement sensed by qubits near a radiation impact.
As TLS scrambling brings TLSs out of or in to resonance with the qubit, the lifetime of the qubit increases or decreases.
Our findings thus identify radiation as a new contribution to fluctuations in qubit lifetimes, with implications for efforts to characterize and improve device stability.

\end{abstract}

\pacs{Valid PACS appear here}
\maketitle

\section{Introduction}

The drive to build a superconducting quantum computer has led to rapid increases in both the number of qubits in a device and the lifetimes, $T_1$, of those qubits. The increase in lifetimes has been remarkable given the sensitivity of the qubits to environmental noise \cite{siddiqi_engineering_2021}. This sensitivity can be harnessed by using the qubits as sensors to better understand the noise. For example, two-level systems (TLSs) are currently a dominant loss mechanism in superconducting qubits, but superconducting qubits are also useful to study TLSs \cite{muller_towards_2019}. A key diagnostic tool has been TLS spectroscopy, in which $T_1$ is measured as the frequency of the qubit is swept \cite{lisenfeld_electric_2019, barends_2013_coherent, bilmes_2021_quantum, meissner_2018_probing, klimov_fluctuations_2018}.
An individual TLS can be resolved as a dip in $T_1$ as the qubit is tuned to the TLS frequency. 
Spectroscopy has revealed that TLSs can drift, appear, and disappear over time \cite{carroll_dynamics_2021, klimov_fluctuations_2018, meissner_2018_probing}.  When a TLS moves in to resonance with a qubit, $T_1$ can be suppressed by an order of magnitude in a modern device \cite{klimov_fluctuations_2018, carroll_dynamics_2021}. This instability in $T_1$ is a threat to quantum computers: when lifetimes decrease, either the computation suffers or the device must be taken offline and retuned, potentially by changing the qubit frequency or by relearning a noise model for error mitigation \cite{berg2022probabilistic}. Therefore understanding TLS dynamics is important to improving superconducting quantum processors. Prior work on TLS dynamics has focused on the interactions between TLSs \cite{faoro_interacting_2015, burnett_2014_evidence}. In this model each high-frequency TLS (i.e. TLS that are near resonant with the qubit frequency) is coupled to many low-frequency TLSs that occasionally flip states due to the ambient thermal energy, perturbing the high-frequency TLS. Evidence for this model has been observed in both superconducting qubits \cite{muller_interacting_2015, klimov_fluctuations_2018, schlor_correlating_2019, burnett_decoherence_2019, meissner_2018_probing} and resonators \cite{burnett_2014_evidence, neill_2013_fluctuations, bejanin_2022_fluctuation, deGraaf_2021_quantifying}. However, TLS-TLS interactions, via either electric or elastic dipole interactions, are very short range \cite{lisenfeld_2015_observation, muller_interacting_2015} compared to the large size of the transmon capacitor paddles ($\sim$100~$\mu$m) or the Josephson junction ($\sim$100~nm), so it is unlikely that multiple high-frequency TLSs would interact with the same set of low-frequency TLSs.  Therefore the interacting defect model would struggle to explain simultaneous dynamics in multiple high-frequency TLSs.

\begin{figure*}[t!]
	\centering
    \includegraphics[width=\textwidth]{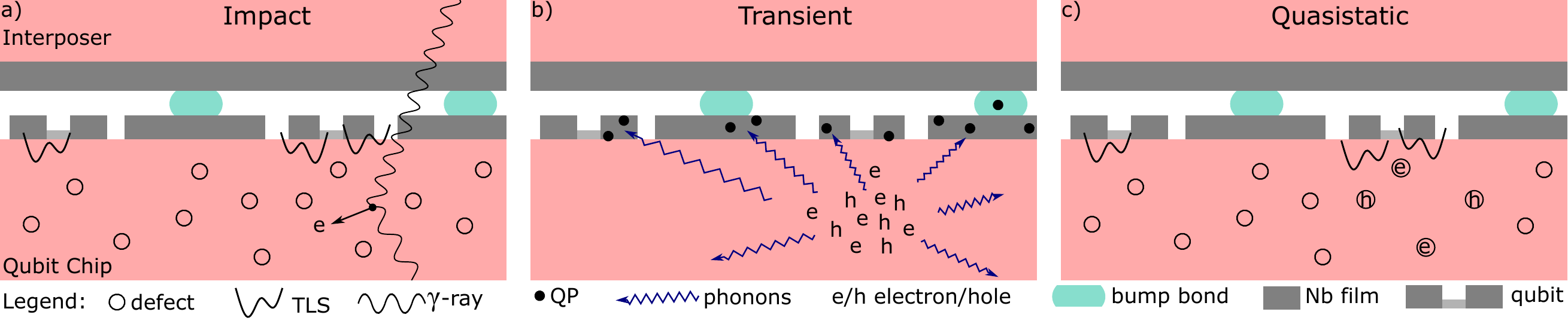}
    \caption{Cartoon of a typical $\gamma$-ray impact. The device consists of a qubit chip bump bonded onto an interposer. Each chip includes a thin, patterned niobium film (gray) on a silicon substrate (pink). a) A $\gamma$-ray Compton scatters off of a silicon atom, ejecting an electron. b) Transient response of the device after the impact. (Defects and TLSs are not shown in b) to avoid clutter.) The electron scatters off of other atoms in the substrate generating electron-hole pairs.  Relaxation and recombination of the electrons and holes creates phonons that propagate outward, potentially for several millimeters, creating QPs when they hit either the ground plane or a qubit.  The bump bonds provide both a lower gap material, which may trap QPs, and a thermalization path for the phonons. c) After the transient response most of the charges recombine, but some of the electrons and holes get trapped at defects, resulting in a long-lasting charge rearrangement, which is sensed by both the local TLS and by the offset-charge on the qubits.}
    \label{fig_cutaway}
\end{figure*}

Improvements in $T_1$  have helped reveal previously unobservable decoherence mechanisms, like the impact of ionizing radiation  \cite{vepsalainen_impact_2020, grunhaupt_loss_2018, cardani_reducing_2021}.  Ionizing radiation generates quasiparticles (QPs) that poison the device, creating a transient dip in $T_1$, but the magnitude and duration of the dip vary widely in the literature \cite{wilen_correlated_2021, mcewen_resolving_2021, vepsalainen_impact_2020, grunhaupt_loss_2018, cardani_reducing_2021}. On multi-qubit devices the QPs can poison many qubits at the same time, potentially the entire chip, generating correlated errors that quantum error correction algorithms struggle to correct \cite{wilen_correlated_2021, mcewen_resolving_2021, martinis_saving_2021, aharonov_2006_fault}. 
Unfortunately background radiation is ubiquitous. $\gamma$-rays and cosmic ray muons cannot be easily shielded, and small amounts of $\alpha$ and $\beta$ radiation sources could even lurk inside the device packaging.  In one recent experiment, background radiation, which hit the chip on average every 10~s, suppressed $T_1$ to less than 1~$\mu$s, and rendered the entire chip unusable for about 100~ms \cite{mcewen_resolving_2021, chen2021exponential}.  Therefore drastic steps have been proposed to mitigate these potentially catastrophic events such as moving the experiments deep underground \cite{cardani_reducing_2021}, adding potentially lossy QP traps \cite{riwar_2016_normal, hosseinkhani_2018_proximity, patel2017phonon, court2008quantitative, iaia_2022_phonon, henriques_2019_phonon, martinis_saving_2021}, and distributing the quantum information across multiple chips \cite{orrell_sensor-assisted_2021, xu_2022_distributed}.

In this paper, we repurpose the standard qubits of a 27-qubit IBM Quantum Falcon R6 processor as multiphysics sensors -- electrometers to detect the impact of radiation and spectrometers to monitor TLS dynamics -- to study the effect of radiation not only on the qubits but also on the TLS. In  section \ref{sec_COJ}, we report localized, multi-qubit offset-charge jumps consistent with the impact of radiation. In contrast to previous reports \cite{wilen_correlated_2021, mcewen_resolving_2021, martinis_saving_2021}, in section \ref{sec_trans_t1} we observed minimal reduction in $T_1$ during impact, showing the potential for superconducting qubits to be robust against ionizing radiation. No special measures were taken to shield the device or to mitigate QPs, suggesting that a combination of materials, packaging, and architecture determines the qubit's susceptibility to radiation.  However, we sometimes observed an unexpected long-lasting change in $T_1$ after the radiation impact. 
In section \ref{sec_TLS}, we show that an impact can scramble the TLS spectrum, by which we mean that several TLSs that are near resonant with the qubit either appear, disappear, or change frequency at the same time.
Because scrambling involves multiple high-frequency TLSs, the interacting defect model is insufficient to explain these dynamics. 
In section \ref{sec_origin} we suggest that the offset-charge jumps and TLS scrambling can both be explained by charge rearrangement after the radiation impact event.
We thus identify a new mechanism contributing to fluctuations in $T_1$ over time, defining a new focus for research to improve stability of superconducting quantum processors.

\section{Charge detection of radiation impact events}\label{sec_COJ}

First we briefly review the dynamics of radiation impacting a superconducting qubit chip as described in the literature \cite{martinis_saving_2021, wilen_correlated_2021, vepsalainen_impact_2020, mcewen_resolving_2021, leman_2012_invited}.  We did not deliberately introduce any radioactive sources \cite{vepsalainen_impact_2020, cardani_reducing_2021}, so we consider only background radiation. Multiple layers of shielding protect our device from low-energy photons such as thermal radiation from the higher temperature stages \cite{gordon_2022_environmental, corcoles_2011_protecting, barends_2011_minimizing}, and also attenuate any $\alpha$ or $\beta$ radiation coming from outside the qubit packaging.  However, $\gamma$-rays and cosmic ray muons or neutrons cannot be easily shielded, and small amounts of $\alpha$ and $\beta$ sources can exist in the qubit packaging. Here we will discuss the impact of a $\gamma$-ray, but the impact of other forms of radiation would be similar. The likely radioisotopes in the vicinity, such as $^{40}$K, $^{232}$Th, and $^{238}$U in building materials like concrete, emit $\gamma$-rays with an energy of order 1~MeV \cite{vepsalainen_impact_2020, wilen_correlated_2021, cardani_reducing_2021}. At these energies, Compton scattering is the most likely interaction of the $\gamma$-ray with the substrate, as shown in Fig. \ref{fig_cutaway}(a). During Compton scattering, the $\gamma$-ray ionizes a silicon atom, ejecting an electron with of order 100~keV of energy \cite{wilen_correlated_2021}. This high-energy electron initiates a cascade process, scattering off of atomic electrons and generating a large number of electron-hole pairs (Fig. \ref{fig_cutaway}(b)). As the electrons and holes relax and recombine they emit photons and phonons. The phonons downconvert in energy until they travel ballistically through the silicon, potentially for several millimeters \cite{patel2017phonon}. When a phonon reaches the superconductor at the surface of the silicon, it can break Cooper pairs, generating QPs, which are a source of loss for the qubits. Because the phonons can travel for millimeters, many qubits may be simultaneously poisoned by the QPs, resulting in correlated energy relaxation events \cite{wilen_correlated_2021, mcewen_resolving_2021, martinis_saving_2021}.  As shown in Fig. \ref{fig_cutaway}(c) the electrons and holes that do not quickly recombine will diffuse with a characteristic trapping length of a few hundred microns until becoming trapped in the substrate or at the surface, resulting in a long-lasting charge rearrangement that can be sensed by nearby qubits.

\begin{figure}[h]
\includegraphics[]{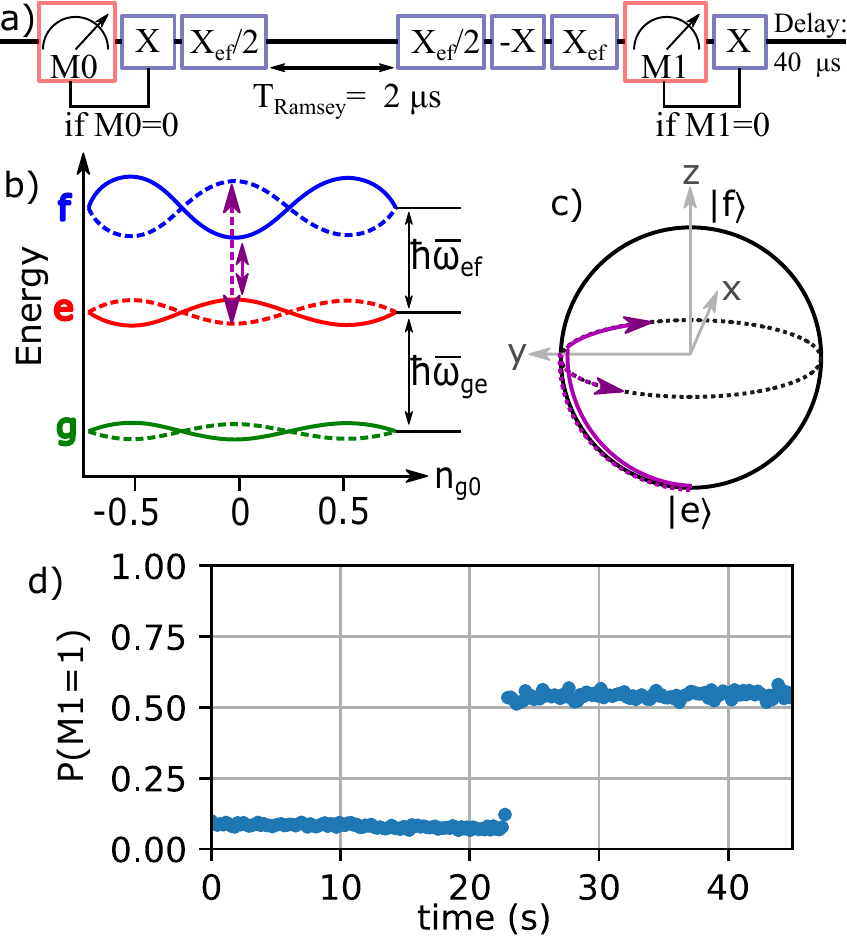}
\caption{Ramsey-based offset-charge jump measurement. a) Measurement conditional $X$ prepares the qubit in $\ket{e}$. $X_{ef}/2-Idle-X_{ef}/2$ performs a fixed delay Ramsey sequence on the $ef$ transition, with an idle time of 2~$\mu$s.  The sequence -$X,X_{ef}$ maps the $ef$ subspace to the $ge$ subspace for Ramsey measurement $M1$. A conditional $X$ then returns the qubit to $\ket{e}$, and a 40~$\mu$s delay between repetitions provides a fixed-delay $T_1$ measurement using the outcome of $M0$.  b) Exaggerated transmon energy level diagram showing the dependence on the unitless and periodic offset-charge, $n_{g0}$. For any $n_{g0}$ there are two $ef$-transition frequencies (purple), one for each QP parity, symmetrically detuned about the mean transition frequency $\bar{\omega}_{ef}$.  c) Bloch sphere illustration of the qubit state evolution during the early part of the Ramsey sequence, showing rotation along the equator at equal rates but in opposite directions depending on the QP parity  (d) A single qubit trace of $P(M1\!=\!1)$ as a function of experiment time, showing a jump 23~s into the experiment. Here the 1 million repetitions were grouped into 200 time bins to compute probabilities.}
\label{fig_ng_jumps}
\end{figure}

Radiation impact events can be detected using different techniques, so we choose based on the capabilities of our quantum processor. The low-$Q$ (Q $\sim$ 1200) resonators used for fast readout ruled out impact detection methods based on changes in kinetic inductance \cite{baselmans_kinetic_2012, grunhaupt_loss_2018, cardani_reducing_2021}. Another method, monitoring for bursts of correlated qubit decays indicative of QP poisoning, has been demonstrated in a similar scale quantum processor\cite{mcewen_resolving_2021}; however, we attempted this method and were not able to resolve any events. Instead, following \cite{wilen_correlated_2021}, we used the qubits to sense the abrupt changes in the local charge environment caused by an impact. The energy levels of the transmon, which retains the charge qubit Hamiltonian, weakly depend on the local charge environment, quantified by the offset-charge, $n_{g0}$. Therefore a change in offset-charge causes a small change in the qubit frequency, which can be detected by a Ramsey measurement. We did not make offset-charge sensitive transmons, so our qubits do not have enough charge dispersion to easily detect an offset-charge jump \cite{wilen_correlated_2021}.   However, the higher energy levels have larger charge dispersion (Fig. \ref{fig_ng_jumps}b), so we used the $ef$ qubit subspace instead \cite{tennant_low_2021}.

The experiment to detect an offset-charge jump is shown in Fig. \ref{fig_ng_jumps}(a).  First, measurement $M0$ followed by a conditional $X$ pulse if the qubit is in $\ket{g}$ prepares the qubit in $\ket{e}$.  An $X_{ef}/2$ pulse rotates the qubit from $\ket{e}$ to the equator of the $ef$ Bloch sphere, after which a fixed-delay Ramsey experiment is performed. Driving the $X_{ef}/2$ pulses at the mean transition frequency $\bar{\omega}_{ef}$ causes the transmon to evolve around the equator of the Bloch sphere during $T_{Ramsey}$ at the detuning $\omega_{ef}(p, n_{g0}) - \bar{\omega}_{ef}$, where $p$ is the QP parity. Because we only want to detect offset-charge jumps, and not QP parity flips, here the two QP parities are symmetrically detuned above and below $\bar{\omega}_{ef}$, so they yield the same measurement result (Fig. \ref{fig_ng_jumps}(b,c)). Similar sequences, replacing the second $X_{ef}/2$ with $Y_{ef}/2$, have been used to measure the QP parity \cite{riste_millisecond_2013, tennant_low_2021}. At the end of the experiment we map the state back to the $ge$ subspace and measure the qubit ($M1$),  where $P(M1\!=\!1)=(1+\cos(\epsilon_{ef}\cos(2\pi n_{g0}) T_{Ramsey}/2))/2$, in the absence of decoherence. Our transmons had $ef$ charge dispersions $\epsilon_{ef}/2\pi \sim 800$ kHz. For fixed $T_{Ramsey}$, offset-charge jumps appear as abrupt jumps in $P(M1\!=\!1)$, as shown in Fig. \ref{fig_ng_jumps}(d).   Because the relationship between $P(M1\!=\!1)$ and $n_{g0}$ is nonlinear and periodic, we cannot determine the magnitude of $\Delta n_{g0}$, meaning we cannot use the size of the jumps on different qubits to predict the location of the impact.  

\begin{figure}[h]
\includegraphics[]{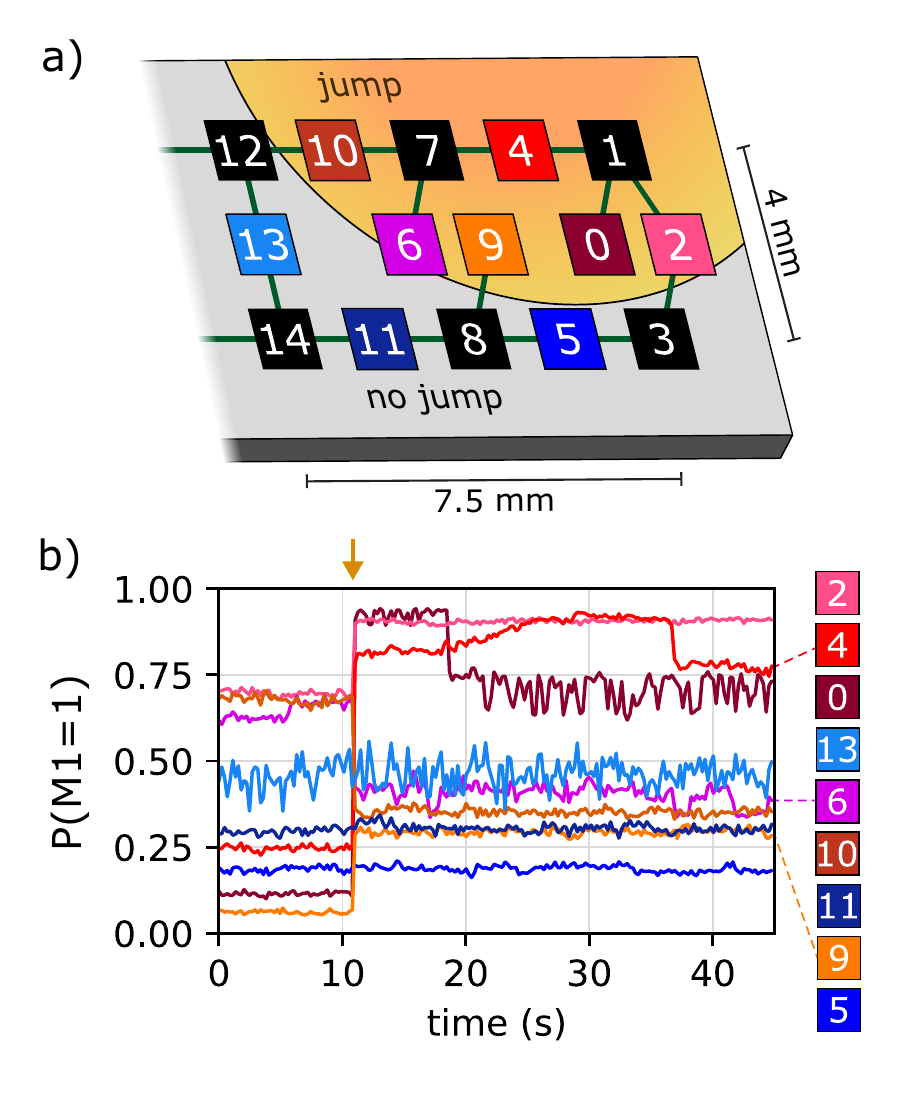}
\caption{Simultaneous offset-charge jumps.  a) A map of the physical layout of half of the device.  The Ramsey based jump detector was run on a set of next nearest neighbors (colored squares), while the other qubits were idled (black squares). b) Results of a jump detection run for 9 qubits on the right half of the device. Simultaneous jumps are observed on 6 out of the 9 qubits, but not on any on the left half of the device (not shown). As seen on the map, the qubits that jumped were localized in the upper right portion of the device (colored gradient), which we attribute to a radiation event impacting that region. } 
\label{fig_big_jump}
\end{figure}

We ran the jump detector simultaneously on qubits across the chip to identify radiation impact events.   Fig. \ref{fig_big_jump}(a) shows the connectivity of half of the qubit chip; the other half is similar. Black lines between qubits indicate couplings via bus resonators that mediate two-qubit gates.  These couplings induce a small shift in the qubit frequency that depends on the state of its neighbors. To avoid contamination of the Ramsey phase by this interaction, we restricted the simultaneous jump detection to a set of 17 non-neighboring qubits (0, 2, 4, 5, 6, 9, 10, 11, 13, 15, 16, 17, 20, 21, 22, 24, 26) while the remaining 10 qubits were idled.  The results from one run of the jump detector are shown in Fig. \ref{fig_big_jump}(b).  About 11~s into the run, simultaneous jumps can be seen on qubits 0, 2, 4, 6, 9, 10 while no jumps were observed on qubits 5, 11, 13 or any of the qubits on the left half of the chip (not shown). The qubits that jumped were localized to a small area of the chip, with a radius of a few millimeters, as shown by the colored background in Fig. \ref{fig_big_jump}(a). Random coincidence cannot explain simultaneous jumps on so many qubits in a small portion of the device (App. \ref{app_algo_multiq}), so these qubits must be sensing a common change in the local environment.  While models of offset-charge drift have traditionally focused on local charge rearrangement very close to an island such as a TLS dipole flip, metallic grain charging or a fluctuating patch potential \cite{stewart_2016_stability, zimmerman2008long, kafanov2008charge, christensen_anomalous_2019, pourkabirian_2014_nonequilibrium}, these models fail to explain simultaneous discrete jumps on qubits that are millimeters apart and well isolated by ground planes (Fig. \ref{fig_cutaway}). We thus attribute these multi-qubit jumps to radiation generating large charge rearrangements in the substrate, which are not as effectively screened by the ground plane. As further evidence that the jumps we observe are due to radiation, in App. \ref{app_comp_rate} we show that the rate of events we extract from our device is consistent with values from the literature. In our device the typical spacing between physically adjacent qubits is 1 to 2~mm, which has been too distant to detect simultaneous jumps in prior experiments \cite{wilen_correlated_2021, pan_2022_engineering}; however, a transmon's effective sensing distance will depend on the geometry and layout of the qubit, the substrate thickness, and the isolation of the qubit from the ground plane, so will vary from device to device.

\section{Transient effect on $T_1$}\label{sec_trans_t1}

\begin{figure}[h]
\includegraphics[width=0.9\columnwidth]{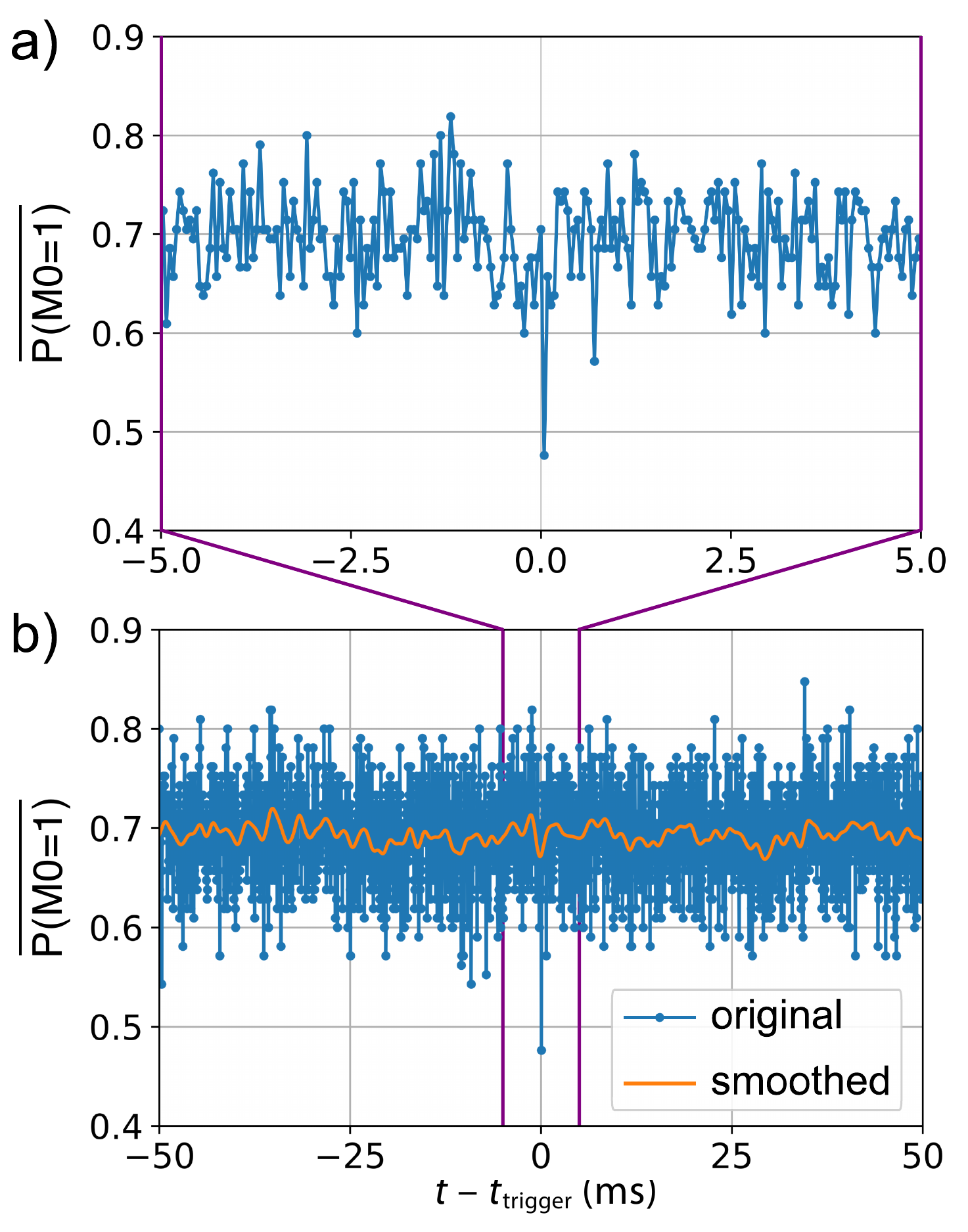}
\caption{Search for excess qubit loss associated with multi-qubit offset-charge jumps. Collecting the 108 constituent single-qubit jumps, we averaged the probability of the qubit remaining in $\ket{e}$ after the 40~$\mu$s delay between shots.  We show $\overline{P(M0\!=\!1)}$ (blue) for (a) 5~ms and (b) 50~ms before and after the jump, based on the jump time extracted from the jump detection algorithm. Smoothing with a Gaussian filter with a sigma of 10 data points (0.44~ms, orange) helps show the absence of a trend. Excess qubit loss due to transient QPs corresponds to a dip in this plot.  We see that $P\overline{(M0\!=\!1)}$ for the data point immediately after the extracted $t_{trigger}$ is suppressed by 4.9~$\sigma$ compared to the background. This is consistent with a small amount of qubit loss induced by a short-lived excess of QPs.}
\label{fig_qp_t1}
\end{figure}

Having demonstrated the detection of radiation impacts, we next looked for a reduction in $T_1$ due to the QPs generated by the impact. The pulse sequence in Fig. \ref{fig_ng_jumps}(a) uses a 40~$\mu$s delay between repetitions as a fixed-delay $T_1$ measurement, acquired in $M0$, interleaved with the jump detection. This sequence was repeated 1 million times per experiment, which lasted 44~s.  We ran this experiment on the above set of non-neighboring qubits 500 times, resulting in a total `detector time' of over 6 hours.  As detailed in App. \ref{app_algo}, we used a matched filter and thresholding to detect multi-qubit jumps and calculate the most likely time step of each jump, $t_{\textrm{trigger}}$.  Using a selective threshold value, we detected 43 multi-qubit jumps comprising a total of 108 single-qubit jumps.  For those 108 events we then looked for excess decays during the fixed-delay $T_1$ measurement.  In Fig. \ref{fig_qp_t1} we plot $\overline{P(M0\!=\!1)}$ for 50~ms before and after the jump, with $t_{\textrm{trigger}}$ aligned for each qubit. The overline is to indicate that data from multiple qubits were averaged together. Excess loss due to transient QP generated during an event will show up as a dip in this plot, with a width set by the timing precision of the jump detection algorithm and the duration of the reduction in $T_1$ \cite{wilen_correlated_2021}.

The averaged data (Fig. \ref{fig_qp_t1}) are well described by a Gaussian distribution centered about a constant background level of $\overline{P(M0=1)}\approx0.7$ (corresponding to an average $T_1\sim100~\mu$s), except for the point immediately after $t=t_{\textrm{trigger}}$ which is 4.9~$\sigma$ below the mean.  Since the nearby data points do not indicate a broader dip, either the data point is somehow spurious, or it suggests that loss occurs only in a very short interval around the impact, e.g. QP-poisoning that recovers within the 44~$\mu$s repetition time. To help assess feasibility of our scheme resolving a real single-point feature,  in App. \ref{app_algo_timing} we analyze simulated data, and find that in some conditions the analysis algorithm can indeed be accurate to within a few time steps. App. \ref{app_algo_thresh} also shows how the depth of the dip depends on the threshold used to declare a jump.

The absence of a sustained dip in Fig. \ref{fig_qp_t1} demonstrates the resilience of our quantum processor to the impact of radiation. We have not taken any special measures to shield or reduce environmental radiation \cite{cardani_reducing_2021, vepsalainen_impact_2020} or added any structures specifically to reduce the impact of transient QPs \cite{martinis_saving_2021, iaia_2022_phonon, karatsu_2019_mitigation, henriques_2019_phonon, riwar_2016_normal, hosseinkhani_2018_proximity, riwar_2019_efficient, pan_2022_engineering}.  In App. \ref{app_algo_multiq} we show that the impact rate that we measure is very similar to previous measurements.  Therefore the robustness of $T_1$ to radiation must derive from the packaging, design,  or materials of the device.  These qubits consist of niobium capacitor pads with aluminum leads for the Josephson junction. The bump bonds provide both a path for thermalization \cite{martinis_saving_2021}, and a lower gap trap for QPs from the niobium ground plane. Going forward additional reports will likely elucidate which variables determine the severity of the impact of radiation.

\section{Effect on TLS\MakeLowercase{s}}\label{sec_TLS}

While looking for a transient hit to $T_1$, we noticed that occasionally the $T_1$ of an individual qubit would suddenly change after a multi-qubit jump. Unlike the transient dips previously discussed, here the $T_1$ would be stable at the new value, which could be higher or lower.  
Next we will show that this is due to radiation ``scrambling'' the TLSs on a qubit, simultaneously changing the frequency of several TLS near resonance with the qubit.  When the scrambling brings a TLS in to (out of) resonance with the qubit, the qubit $T_1$ decreases (increases).  After the scrambling, the TLS spectrum is stable, so $T_1$ is stable at the new value.

\begin{figure}[h]
\includegraphics[]{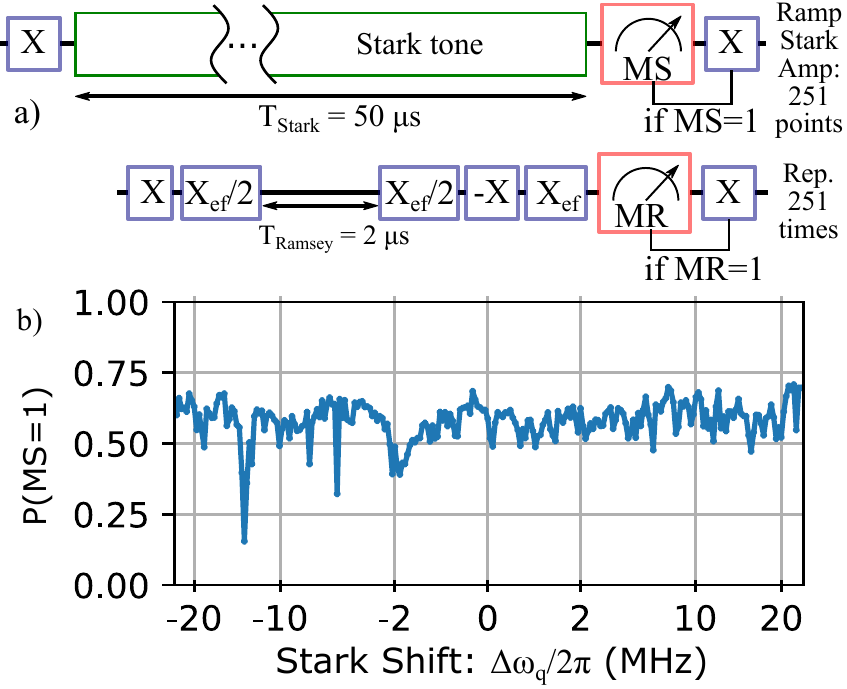}
\caption{Interleaved TLS spectroscopy and jump detection. a) Each iteration, TLS spectroscopy (upper sequence) was executed by ramping the amplitude of the Stark pulse to sweep the qubit frequency, before being measured during $MS$; then the jump detector (lower sequence) was repeated 251 times, before being measured during $MR$. A conditional $X$ to reset the qubit and a 20~$\mu$s delay followed each measurement. Each 4 minute run consisted of 10,000 iterations of this sequence. b) A TLS spectrum, showing $P(MS\!=\!1)$ as a function of the qubit Stark shift, averaged over all 10,000 iterations.  Dips, such as that at $\Delta \omega_q$ = -14~MHz, correspond to excess loss when the qubit is in resonance with a TLS.}
\label{fig_stark_TLS}
\end{figure}

We used TLS spectroscopy to monitor TLSs near resonance with the qubit.  During TLS spectroscopy the qubit frequency is swept, and when the qubit is brought in to resonance with a TLS, the two hybridize, decreasing the qubit $T_1$ \cite{klimov_fluctuations_2018, lisenfeld_electric_2019, barends_2013_coherent, bilmes_2021_quantum, meissner_2018_probing}. Our transmons are not flux-tunable, but can be tuned via the Stark shift to probe TLSs using the pulse sequence shown in Fig. \ref{fig_stark_TLS}(a) \cite{carroll_dynamics_2021,  zhao_2022_combating}.  A Stark tone amplitude $\Omega_s$ shifts the qubit frequency by  
$ \Delta \omega_q = \frac{\alpha \Omega_s^2}{2\Delta_{s}\left( \alpha + \Delta_s \right) } $,
where $\alpha$ is the qubit anharmonicity and $\Delta_s$ is the detuning of the Stark tone from the unshifted qubit frequency. With fixed detuning ($\Delta_s$ = $\pm$~50~MHz) we stepped $\Omega_s$ to sweep the Stark shifted qubit frequency from $\Delta \omega_q \approx$ -20~MHz to +20~MHz. Fig. \ref{fig_stark_TLS}(b) shows one example of $P(MS\!=\!1)$ as a function of $\Delta \omega_q$. We linearly ramped $\Omega_s$, resulting in the quadratic spacing of $\Delta \omega_q$ in the horizontal axis.  Dips in $P(MS\!=\!1)$ indicate excess loss attributable to TLSs. To look for changes in the TLS spectroscopy coincident with multi-qubit jumps, we interleaved the TLS spectroscopy with the jump detector (Fig. \ref{fig_stark_TLS}(a)). Each iteration consisted of two parts: TLS spectroscopy in which the amplitude of the Stark pulse was swept in 251 steps, and 251 repetitions of the jump detector.  A conditional $X$ followed by a 20~$\mu$s delay prepared the qubit for the next repetition. This 24 ms sequence was repeated 10,000 times, taking a total of 4 minutes. We repeated the experiment 172 times for over 11 hours of `detector time'.  A modified jump detection algorithm (App. \ref{app_algo_scramble}) run on the Ramsey jump sequence identified 656 multi-qubit jumps. Interleaving the jump detector and the TLS spectroscopy allowed us to look for changes in the TLS spectra associated with these multi-qubit jumps, and thus look for the impact of radiation on the TLSs in the device.

\begin{figure*}[t!]
	\centering
    \includegraphics[]{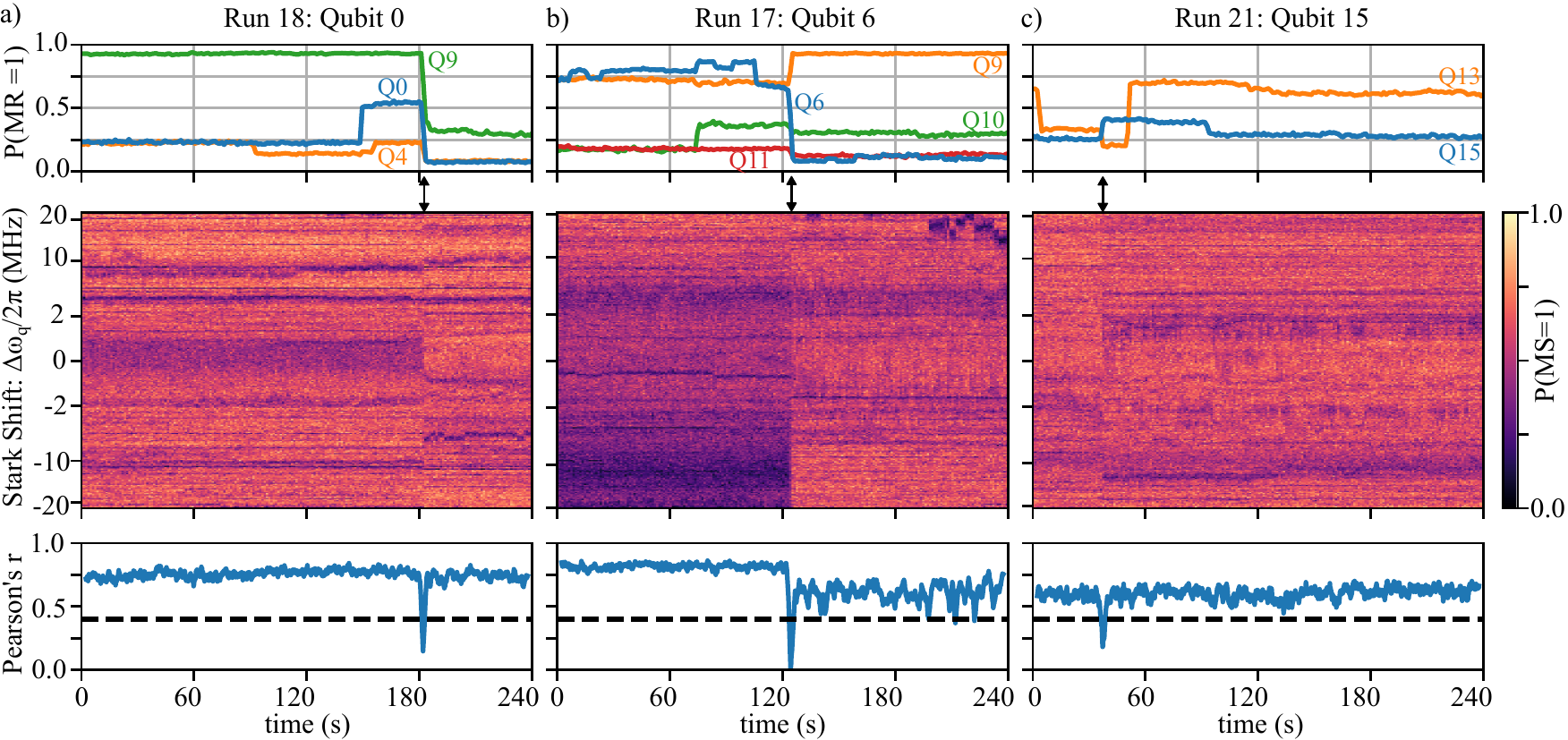}
    \caption{TLS scrambling due to radiation.  Each column (a-c) shows a TLS scrambling event from a different run of the experiment. The time axis is shared across all panels in a column. Values in the top and middle rows have been averaged within 200 time bins to compute probabilities. Top panels show outcomes of Ramsey jump detection ($MR$). Middle panels show evolution of TLS spectra over time. The frequency scale is nonlinear because of the quadratic dependence on Stark tone amplitude. The color scale represents $P(MS\!=\!1)$ at the end of the 50~$\mu$s Stark tone; dark horizontal lines correspond to excess loss when the qubit is resonant with a TLS.  Bottom panels shows Pearson's $r$ for the TLS spectroscopy as defined in the main text.   Dips correspond to sudden, pronounced changes in spectral features in the middle panel. Dips crossing the threshold $r$ = 0.4 (black dashed line) simultaneous with a multi-qubit jump were declared TLS scrambling events.  Because we previously established multi-qubit jumps are caused by the impact of radiation, we argue that the simultaneous TLS scrambling is also caused by the impact of radiation. } 
    \label{fig_stark_3_events}
\end{figure*}

The three columns of Fig. \ref{fig_stark_3_events} show three examples of TLS scrambling, each coincident with a multi-qubit jump.  In the top panel of Fig. \ref{fig_stark_3_events}(a) we plot the output of the jump detector for one run on qubits 0, 4, and 9, all of which experience a simultaneous jump 180~s into the experiment, indicating a radiation impact event. Interleaved TLS spectroscopy (middle panel) shows that the event coincides with abrupt changes in the spectral signatures of multiple TLSs (dark horizontal bands). The TLSs are stable both before and after the event. The dynamics during scrambling are too fast to resolve in this experiment. Notice that prior to scrambling, when there is no Stark shift ($\Omega_S = 0$), $T_1$ suffers due to a TLS near resonance with the qubit.  The scrambling brings that TLS out of resonance, resulting in a long-lasting change to $T_1$. The other two columns in Fig. \ref{fig_stark_3_events} show two additional examples of TLS scrambling events.

The jump detector traces in the top row of Fig. \ref{fig_stark_3_events} show additional jumps that do not correspond to TLS scrambling.  In fact, most offset-charge jumps are not accompanied by TLS scrambling.  To determine the fraction of multi-qubit jumps associated with a TLS scrambling event, we used the Pearson correlation coefficient $r$ to quantify changes in the TLS spectra as shown in the bottom row of Fig. \ref{fig_stark_3_events}.  Pearson's $r$ measures the linear correlation between two ordered data sets, $x$ and $y$, and is defined as the dot product $r = \hat{x}\cdot\hat{y}$ of the normalized, zero-mean sets $\hat{x}=(x-\bar{x})/\sigma_x$, where $\sigma_x$ is the standard deviation of $x$. 
For each time step, $x(y)$ was the average $P(MS\!=\!1)$ for 200 data points, 4.8~s, before (after) the time step. Pearson's $r$ is 1(0) if the two data sets are perfectly (un)correlated. In the bottom panels of Fig. \ref{fig_stark_3_events}, we see the TLS scrambling events weaken the correlations between past and future spectra, causing $r$ to dip below our selection threshold of 0.4 (black dashed line). However, crossing this threshold did not guarantee a scrambling event.  For example, in the final minute of Fig. \ref{fig_stark_3_events}(b), a single pronounced TLS undergoing diffusion near $\Delta_S$ = 20~MHz, presumably due to interaction with nearby thermal fluctuators, caused $r$ to dip below 0.4 without TLS scrambling. Thus we observe both single TLS diffusion due to thermal-fluctuators and TLS scrambling which affects multiple TLSs simultaneously. 

Of all $r$ values computed for each qubit at each time-point, $r$ only fell below the 0.4 threshold $0.06\%$ of the time.  In contrast, one of the qubits participating in a multi-qubit jump had an $r$ below 0.4 during 34 of the 656 multi-qubit jumps (5.2\%).  Therefore, because both multi-qubit jumps and events where $r$ dips below 0.4 are rare, it is unlikely that we would observe both simultaneously without a common origin.  Having established that radiation is the most likely origin of the multi-qubit jumps, we infer that radiation is also the likely origin of the TLS scrambling.

Most of the detected radiation impact events, as witnessed by multi-qubit jumps, were not accompanied by detectable TLS scrambling. Also in this data set we observed an absence of TLS scrambling on multiple qubits during the same event. This suggests a picture in which the offset-charge on one of our qubits can sense a distant impact event, while the TLSs are less sensitive to distant events. We conjecture that some of the multi-qubit jumps without TLS scrambling would be due to impacts that were not close enough to a qubit to induce TLS scrambling. Although this is a qualitatively plausible picture, we highlight that it is not the only possible explanation for this difference in response. There could be a higher energy threshold for the impact to scramble TLSs beyond the energy required to cause a multi-qubit jump.  Alternatively, TLS scrambling might be associated with only certain types of radiation.

\section{Discussion}\label{sec_origin}
What can we say about the interaction mechanism underlying TLS scrambling? Although the physical origin of TLSs remains the subject of much debate \cite{muller_towards_2019}, TLSs are known to couple to the local electric field via a dipole moment. Given their simultaneity, it is natural to suppose the TLS scrambling and offset-charge jumps are both responses to the same redistribution of charge. As shown in the insets of Fig. \ref{fig_cutaway}, some of the electrons and holes generated during the impact escape recombination and diffuse until becoming trapped at defects.  This charge redistribution will change the electric field at the TLS, thus changing its frequency. The dipole moment of the TLS ($\sim$ 1 e\AA \cite{muller_towards_2019}) is much smaller than the transmon dipole moment ($\sim$ 100~e$\mu$m \cite{paik_2011_observation}), potentially explaining why qubits over a large area participate in a jump, but the TLS scrambling is localized to the TLS in the vicinity of at most one qubit. Because this picture predicts some TLSs will scramble more than others, with TLSs at the metal-substrate or substrate-air interfaces more sensitive to charge rearrangements in the substrate than TLSs in the junction or at the metal-air interface, further insight could be gained by monitoring scrambling in a setup able to locate individual TLS within a device \cite{bilmes_2020_resolving}.

Other mechanisms could also be at play. The scrambling could be mediated by the TLS elastic dipole coupling to local strains. Heat generated by an impact could cause brief expansion and thus local strain shifting the TLS \cite{nazaretski_2004_effect}, akin to thermal cycling of the cryostat. Recently stress-induced micro-fractures have been suggested as an alternative source of phonon and QP bursts\cite{anthony2022stress}, but in our architecture the electrical fields from micro-fractures due to the stress at the bump-bonds or niobium-silicon interface will be screened by the ground plane, thus unlikely to cause the offset-charge jumps. Repeating these measurements in different processors could reveal whether these dynamics depend on the materials and architecture, as the response to transient QPs appears to. It would also be enlightening to study other types of qubits or similarly susceptible devices such as single-electron transistors and SQUIDs \cite{stewart_2016_stability, gusenkova2022operating}.

Many proposed explanations for TLSs involve trapped charges: tunneling electrons \cite{faoro_2006_quantum, agarwal_2013_polaronic, lutchyn_2008_quantum}, localized metal-induced gap states at the metal-insulator interface \cite{choi_2009_localization, pourkabirian_2014_nonequilibrium}, or trapped QPs \cite{bespalov_2016_theoretical, deGraaf_2020_qpTLS}. Because these charges can be created and destroyed during the impact, investigations of the response of TLSs to the impact of radiation may also tell us about the nature of the TLSs themselves.

\section{Conclusion}\label{sec_conc}

Using the standard transmons on a quantum processor as an array of electrometers, we detected radiation impacts and looked for associated effects on $T_1$. We were barely able to resolve any transient effect, demonstrating a robustness to radiation in contrast with the severe hit to $T_1$ in the literature  \cite{wilen_correlated_2021, mcewen_resolving_2021, martinis_saving_2021}. We did not take any steps to reduce the rate or severity of the impacts, suggesting the resilience derives from the device materials and architecture, and moreover can be obtained without compromising the baseline $T_1$. These results suggest that one may hope to realize traditional quantum error correction without the drastic interventions suggested in the literature \cite{cardani_reducing_2021, riwar_2016_normal, hosseinkhani_2018_proximity, patel2017phonon, court2008quantitative, iaia_2022_phonon, henriques_2019_phonon, martinis_saving_2021, orrell_sensor-assisted_2021, xu_2022_distributed} 

Some radiation impacts coincided with long-lasting changes in $T_1$, both positive and negative, which we explained through radiation-induced TLS scrambling. These results revealed a new type of TLS dynamics beyond the interacting defect model. We therefore identify radiation as one driver of fluctuations in $T_1$ over time, which is a central challenge to scalable superconducting quantum computing. Looking to the future, one can even imagine using an ionizing radiation source to controllably reset an uncooperative TLS spectrum plaguing a large quantum processor, with no need for a disruptive thermal cycle of the cryostat.

\section{Acknowledgements}

The device was designed and fabricated internally at IBM. We acknowledge the use of IBM Quantum services for this work, and these results were enabled by the work of the IBM Quantum software and hardware teams. This work was supported by IARPA under LogiQ (contract W911NF-16-1-0114). All statements of fact, opinion or conclusions contained herein are those of the authors and should not be construed as representing the official views or policies of the US Government.
We thank  Oliver Dial, Francesco Valenti, Youngseok Kim, Sami Rosenblatt, Joey Suttle, Jared Hertzberg, Chris Lirakis, Luke Govia, Ken Rodbell,  Antonio C{\'o}rcoles, Maika Takita and Matthias Steffen for helpful conversations, device bring-up,  and programmatic support.


\begin{thebibliography}{61}%
	\makeatletter
	\providecommand \@ifxundefined [1]{%
		\@ifx{#1\undefined}
	}%
	\providecommand \@ifnum [1]{%
		\ifnum #1\expandafter \@firstoftwo
		\else \expandafter \@secondoftwo
		\fi
	}%
	\providecommand \@ifx [1]{%
		\ifx #1\expandafter \@firstoftwo
		\else \expandafter \@secondoftwo
		\fi
	}%
	\providecommand \natexlab [1]{#1}%
	\providecommand \enquote  [1]{``#1''}%
	\providecommand \bibnamefont  [1]{#1}%
	\providecommand \bibfnamefont [1]{#1}%
	\providecommand \citenamefont [1]{#1}%
	\providecommand \href@noop [0]{\@secondoftwo}%
	\providecommand \href [0]{\begingroup \@sanitize@url \@href}%
	\providecommand \@href[1]{\@@startlink{#1}\@@href}%
	\providecommand \@@href[1]{\endgroup#1\@@endlink}%
	\providecommand \@sanitize@url [0]{\catcode `\\12\catcode `\$12\catcode
		`\&12\catcode `\#12\catcode `\^12\catcode `\_12\catcode `\%12\relax}%
	\providecommand \@@startlink[1]{}%
	\providecommand \@@endlink[0]{}%
	\providecommand \url  [0]{\begingroup\@sanitize@url \@url }%
	\providecommand \@url [1]{\endgroup\@href {#1}{\urlprefix }}%
	\providecommand \urlprefix  [0]{URL }%
	\providecommand \Eprint [0]{\href }%
	\providecommand \doibase [0]{https://doi.org/}%
	\providecommand \selectlanguage [0]{\@gobble}%
	\providecommand \bibinfo  [0]{\@secondoftwo}%
	\providecommand \bibfield  [0]{\@secondoftwo}%
	\providecommand \translation [1]{[#1]}%
	\providecommand \BibitemOpen [0]{}%
	\providecommand \bibitemStop [0]{}%
	\providecommand \bibitemNoStop [0]{.\EOS\space}%
	\providecommand \EOS [0]{\spacefactor3000\relax}%
	\providecommand \BibitemShut  [1]{\csname bibitem#1\endcsname}%
	\let\auto@bib@innerbib\@empty
	\bibitem [{\citenamefont {Siddiqi}(2021)}]{siddiqi_engineering_2021}%
	\BibitemOpen
	\bibfield  {author} {\bibinfo {author} {\bibfnamefont {I.}~\bibnamefont
			{Siddiqi}},\ }\bibfield  {title} {\bibinfo {title} {Engineering
			high-coherence superconducting qubits},\ }\href@noop {} {\bibfield  {journal}
		{\bibinfo  {journal} {Nat. Rev. Mat.}\ }\textbf {\bibinfo {volume} {6}}
		(\bibinfo {year} {2021})}\BibitemShut {NoStop}%
	\bibitem [{\citenamefont {Müller}\ \emph {et~al.}(2019)\citenamefont
		{Müller}, \citenamefont {Cole},\ and\ \citenamefont
		{Lisenfeld}}]{muller_towards_2019}%
	\BibitemOpen
	\bibfield  {author} {\bibinfo {author} {\bibfnamefont {C.}~\bibnamefont
			{Müller}}, \bibinfo {author} {\bibfnamefont {J.~H.}\ \bibnamefont {Cole}},\
		and\ \bibinfo {author} {\bibfnamefont {J.}~\bibnamefont {Lisenfeld}},\
	}\bibfield  {title} {\bibinfo {title} {Towards understanding
			two-level-systems in amorphous solids: insights from quantum circuits},\
	}\href@noop {} {\bibfield  {journal} {\bibinfo  {journal} {Rep. Prog. Phys.}\
		}\textbf {\bibinfo {volume} {82}},\ \bibinfo {pages} {124501} (\bibinfo
		{year} {2019})}\BibitemShut {NoStop}%
	\bibitem [{\citenamefont {Lisenfeld}\ \emph {et~al.}(2019)\citenamefont
		{Lisenfeld}, \citenamefont {Bilmes}, \citenamefont {Megrant}, \citenamefont
		{Barends}, \citenamefont {Kelly}, \citenamefont {Klimov}, \citenamefont
		{Weiss}, \citenamefont {Martinis},\ and\ \citenamefont
		{Ustinov}}]{lisenfeld_electric_2019}%
	\BibitemOpen
	\bibfield  {author} {\bibinfo {author} {\bibfnamefont {J.}~\bibnamefont
			{Lisenfeld}}, \bibinfo {author} {\bibfnamefont {A.}~\bibnamefont {Bilmes}},
		\bibinfo {author} {\bibfnamefont {A.}~\bibnamefont {Megrant}}, \bibinfo
		{author} {\bibfnamefont {R.}~\bibnamefont {Barends}}, \bibinfo {author}
		{\bibfnamefont {J.}~\bibnamefont {Kelly}}, \bibinfo {author} {\bibfnamefont
			{P.}~\bibnamefont {Klimov}}, \bibinfo {author} {\bibfnamefont
			{G.}~\bibnamefont {Weiss}}, \bibinfo {author} {\bibfnamefont {J.~M.}\
			\bibnamefont {Martinis}},\ and\ \bibinfo {author} {\bibfnamefont {A.~V.}\
			\bibnamefont {Ustinov}},\ }\bibfield  {title} {\bibinfo {title} {Electric
			field spectroscopy of material defects in transmon qubits},\ }\href@noop {}
	{\bibfield  {journal} {\bibinfo  {journal} {npj Quantum Inf.}\ }\textbf
		{\bibinfo {volume} {5}} (\bibinfo {year} {2019})}\BibitemShut {NoStop}%
	\bibitem [{\citenamefont {Barends}\ \emph {et~al.}(2013)\citenamefont
		{Barends}, \citenamefont {Kelly}, \citenamefont {Megrant}, \citenamefont
		{Sank}, \citenamefont {Jeffrey}, \citenamefont {Chen}, \citenamefont {Yin},
		\citenamefont {Chiaro}, \citenamefont {Mutus}, \citenamefont {Neill} \emph
		{et~al.}}]{barends_2013_coherent}%
	\BibitemOpen
	\bibfield  {author} {\bibinfo {author} {\bibfnamefont {R.}~\bibnamefont
			{Barends}}, \bibinfo {author} {\bibfnamefont {J.}~\bibnamefont {Kelly}},
		\bibinfo {author} {\bibfnamefont {A.}~\bibnamefont {Megrant}}, \bibinfo
		{author} {\bibfnamefont {D.}~\bibnamefont {Sank}}, \bibinfo {author}
		{\bibfnamefont {E.}~\bibnamefont {Jeffrey}}, \bibinfo {author} {\bibfnamefont
			{Y.}~\bibnamefont {Chen}}, \bibinfo {author} {\bibfnamefont {Y.}~\bibnamefont
			{Yin}}, \bibinfo {author} {\bibfnamefont {B.}~\bibnamefont {Chiaro}},
		\bibinfo {author} {\bibfnamefont {J.}~\bibnamefont {Mutus}}, \bibinfo
		{author} {\bibfnamefont {C.}~\bibnamefont {Neill}}, \emph {et~al.},\
	}\bibfield  {title} {\bibinfo {title} {Coherent josephson qubit suitable for
			scalable quantum integrated circuits},\ }\href@noop {} {\bibfield  {journal}
		{\bibinfo  {journal} {Phys. Rev. Lett.}\ }\textbf {\bibinfo {volume} {111}},\
		\bibinfo {pages} {080502} (\bibinfo {year} {2013})}\BibitemShut {NoStop}%
	\bibitem [{\citenamefont {Bilmes}\ \emph {et~al.}(2021)\citenamefont {Bilmes},
		\citenamefont {Volosheniuk}, \citenamefont {Brehm}, \citenamefont {Ustinov},\
		and\ \citenamefont {Lisenfeld}}]{bilmes_2021_quantum}%
	\BibitemOpen
	\bibfield  {author} {\bibinfo {author} {\bibfnamefont {A.}~\bibnamefont
			{Bilmes}}, \bibinfo {author} {\bibfnamefont {S.}~\bibnamefont {Volosheniuk}},
		\bibinfo {author} {\bibfnamefont {J.~D.}\ \bibnamefont {Brehm}}, \bibinfo
		{author} {\bibfnamefont {A.~V.}\ \bibnamefont {Ustinov}},\ and\ \bibinfo
		{author} {\bibfnamefont {J.}~\bibnamefont {Lisenfeld}},\ }\bibfield  {title}
	{\bibinfo {title} {Quantum sensors for microscopic tunneling systems},\
	}\href@noop {} {\bibfield  {journal} {\bibinfo  {journal} {npj Quantum Inf.}\
		}\textbf {\bibinfo {volume} {7}},\ \bibinfo {pages} {1} (\bibinfo {year}
		{2021})}\BibitemShut {NoStop}%
	\bibitem [{\citenamefont {Mei{\ss}ner}\ \emph {et~al.}(2018)\citenamefont
		{Mei{\ss}ner}, \citenamefont {Seiler}, \citenamefont {Lisenfeld},
		\citenamefont {Ustinov},\ and\ \citenamefont
		{Weiss}}]{meissner_2018_probing}%
	\BibitemOpen
	\bibfield  {author} {\bibinfo {author} {\bibfnamefont {S.~M.}\ \bibnamefont
			{Mei{\ss}ner}}, \bibinfo {author} {\bibfnamefont {A.}~\bibnamefont {Seiler}},
		\bibinfo {author} {\bibfnamefont {J.}~\bibnamefont {Lisenfeld}}, \bibinfo
		{author} {\bibfnamefont {A.~V.}\ \bibnamefont {Ustinov}},\ and\ \bibinfo
		{author} {\bibfnamefont {G.}~\bibnamefont {Weiss}},\ }\bibfield  {title}
	{\bibinfo {title} {Probing individual tunneling fluctuators with coherently
			controlled tunneling systems},\ }\href@noop {} {\bibfield  {journal}
		{\bibinfo  {journal} {Phys. Rev. B}\ }\textbf {\bibinfo {volume} {97}},\
		\bibinfo {pages} {180505} (\bibinfo {year} {2018})}\BibitemShut {NoStop}%
	\bibitem [{\citenamefont {Klimov}\ \emph {et~al.}(2018)\citenamefont {Klimov},
		\citenamefont {Kelly}, \citenamefont {Chen}, \citenamefont {Neeley},
		\citenamefont {Megrant}, \citenamefont {Burkett}, \citenamefont {Barends},
		\citenamefont {Arya}, \citenamefont {Chiaro}, \citenamefont {Chen} \emph
		{et~al.}}]{klimov_fluctuations_2018}%
	\BibitemOpen
	\bibfield  {author} {\bibinfo {author} {\bibfnamefont {P.}~\bibnamefont
			{Klimov}}, \bibinfo {author} {\bibfnamefont {J.}~\bibnamefont {Kelly}},
		\bibinfo {author} {\bibfnamefont {Z.}~\bibnamefont {Chen}}, \bibinfo {author}
		{\bibfnamefont {M.}~\bibnamefont {Neeley}}, \bibinfo {author} {\bibfnamefont
			{A.}~\bibnamefont {Megrant}}, \bibinfo {author} {\bibfnamefont
			{B.}~\bibnamefont {Burkett}}, \bibinfo {author} {\bibfnamefont
			{R.}~\bibnamefont {Barends}}, \bibinfo {author} {\bibfnamefont
			{K.}~\bibnamefont {Arya}}, \bibinfo {author} {\bibfnamefont {B.}~\bibnamefont
			{Chiaro}}, \bibinfo {author} {\bibfnamefont {Y.}~\bibnamefont {Chen}}, \emph
		{et~al.},\ }\bibfield  {title} {\bibinfo {title} {Fluctuations of
			energy-relaxation times in superconducting qubits},\ }\href
	{https://doi.org/10.1103/PhysRevLett.121.090502} {\bibfield  {journal}
		{\bibinfo  {journal} {Phys. Rev. Lett.}\ }\textbf {\bibinfo {volume} {121}},\
		\bibinfo {pages} {090502} (\bibinfo {year} {2018})}\BibitemShut {NoStop}%
	\bibitem [{\citenamefont {Carroll}\ \emph {et~al.}(2021)\citenamefont
		{Carroll}, \citenamefont {Rosenblatt}, \citenamefont {Jurcevic},
		\citenamefont {Lauer},\ and\ \citenamefont
		{Kandala}}]{carroll_dynamics_2021}%
	\BibitemOpen
	\bibfield  {author} {\bibinfo {author} {\bibfnamefont {M.}~\bibnamefont
			{Carroll}}, \bibinfo {author} {\bibfnamefont {S.}~\bibnamefont {Rosenblatt}},
		\bibinfo {author} {\bibfnamefont {P.}~\bibnamefont {Jurcevic}}, \bibinfo
		{author} {\bibfnamefont {I.}~\bibnamefont {Lauer}},\ and\ \bibinfo {author}
		{\bibfnamefont {A.}~\bibnamefont {Kandala}},\ }\bibfield  {title} {\bibinfo
		{title} {Dynamics of superconducting qubit relaxation times},\ }\href@noop {}
	{\bibfield  {journal} {\bibinfo  {journal} {arXiv:2105.15201}\ } (\bibinfo
		{year} {2021})}\BibitemShut {NoStop}%
	\bibitem [{\citenamefont {van~den Berg}\ \emph {et~al.}(2022)\citenamefont
		{van~den Berg}, \citenamefont {Minev}, \citenamefont {Kandala},\ and\
		\citenamefont {Temme}}]{berg2022probabilistic}%
	\BibitemOpen
	\bibfield  {author} {\bibinfo {author} {\bibfnamefont {E.}~\bibnamefont
			{van~den Berg}}, \bibinfo {author} {\bibfnamefont {Z.~K.}\ \bibnamefont
			{Minev}}, \bibinfo {author} {\bibfnamefont {A.}~\bibnamefont {Kandala}},\
		and\ \bibinfo {author} {\bibfnamefont {K.}~\bibnamefont {Temme}},\ }\bibfield
	{title} {\bibinfo {title} {Probabilistic error cancellation with sparse
			pauli-lindblad models on noisy quantum processors},\ }\href@noop {}
	{\bibfield  {journal} {\bibinfo  {journal} {arXiv:2201.09866}\ } (\bibinfo
		{year} {2022})}\BibitemShut {NoStop}%
	\bibitem [{\citenamefont {Faoro}\ and\ \citenamefont
		{Ioffe}(2015)}]{faoro_interacting_2015}%
	\BibitemOpen
	\bibfield  {author} {\bibinfo {author} {\bibfnamefont {L.}~\bibnamefont
			{Faoro}}\ and\ \bibinfo {author} {\bibfnamefont {L.~B.}\ \bibnamefont
			{Ioffe}},\ }\bibfield  {title} {\bibinfo {title} {Interacting tunneling model
			for two-level systems in amorphous materials and its predictions for their
			dephasing and noise in superconducting microresonators},\ }\href
	{https://doi.org/10.1103/PhysRevB.91.014201} {\bibfield  {journal} {\bibinfo
			{journal} {Phys. Rev. B}\ }\textbf {\bibinfo {volume} {91}},\ \bibinfo
		{pages} {014201} (\bibinfo {year} {2015})}\BibitemShut {NoStop}%
	\bibitem [{\citenamefont {Burnett}\ \emph {et~al.}(2014)\citenamefont
		{Burnett}, \citenamefont {Faoro}, \citenamefont {Wisby}, \citenamefont
		{Gurtovoi}, \citenamefont {Chernykh}, \citenamefont {Mikhailov},
		\citenamefont {Tulin}, \citenamefont {Shaikhaidarov}, \citenamefont
		{Antonov}, \citenamefont {Meeson} \emph {et~al.}}]{burnett_2014_evidence}%
	\BibitemOpen
	\bibfield  {author} {\bibinfo {author} {\bibfnamefont {J.}~\bibnamefont
			{Burnett}}, \bibinfo {author} {\bibfnamefont {L.}~\bibnamefont {Faoro}},
		\bibinfo {author} {\bibfnamefont {I.}~\bibnamefont {Wisby}}, \bibinfo
		{author} {\bibfnamefont {V.}~\bibnamefont {Gurtovoi}}, \bibinfo {author}
		{\bibfnamefont {A.}~\bibnamefont {Chernykh}}, \bibinfo {author}
		{\bibfnamefont {G.}~\bibnamefont {Mikhailov}}, \bibinfo {author}
		{\bibfnamefont {V.}~\bibnamefont {Tulin}}, \bibinfo {author} {\bibfnamefont
			{R.}~\bibnamefont {Shaikhaidarov}}, \bibinfo {author} {\bibfnamefont
			{V.}~\bibnamefont {Antonov}}, \bibinfo {author} {\bibfnamefont
			{P.}~\bibnamefont {Meeson}}, \emph {et~al.},\ }\bibfield  {title} {\bibinfo
		{title} {Evidence for interacting two-level systems from the 1/f noise of a
			superconducting resonator},\ }\href@noop {} {\bibfield  {journal} {\bibinfo
			{journal} {Nat. Commun.}\ }\textbf {\bibinfo {volume} {5}},\ \bibinfo {pages}
		{1} (\bibinfo {year} {2014})}\BibitemShut {NoStop}%
	\bibitem [{\citenamefont {Müller}\ \emph {et~al.}(2015)\citenamefont
		{Müller}, \citenamefont {Lisenfeld}, \citenamefont {Shnirman},\ and\
		\citenamefont {Poletto}}]{muller_interacting_2015}%
	\BibitemOpen
	\bibfield  {author} {\bibinfo {author} {\bibfnamefont {C.}~\bibnamefont
			{Müller}}, \bibinfo {author} {\bibfnamefont {J.}~\bibnamefont {Lisenfeld}},
		\bibinfo {author} {\bibfnamefont {A.}~\bibnamefont {Shnirman}},\ and\
		\bibinfo {author} {\bibfnamefont {S.}~\bibnamefont {Poletto}},\ }\bibfield
	{title} {\bibinfo {title} {Interacting two-level defects as sources of
			fluctuating high-frequency noise in superconducting circuits},\ }\href
	{https://doi.org/10.1103/PhysRevB.92.035442} {\bibfield  {journal} {\bibinfo
			{journal} {Phys. Rev. B}\ }\textbf {\bibinfo {volume} {92}},\ \bibinfo
		{pages} {035442} (\bibinfo {year} {2015})}\BibitemShut {NoStop}%
	\bibitem [{\citenamefont {Schlör}\ \emph {et~al.}(2019)\citenamefont
		{Schlör}, \citenamefont {Lisenfeld}, \citenamefont {Müller}, \citenamefont
		{Bilmes}, \citenamefont {Schneider}, \citenamefont {Pappas}, \citenamefont
		{Ustinov},\ and\ \citenamefont {Weides}}]{schlor_correlating_2019}%
	\BibitemOpen
	\bibfield  {author} {\bibinfo {author} {\bibfnamefont {S.}~\bibnamefont
			{Schlör}}, \bibinfo {author} {\bibfnamefont {J.}~\bibnamefont {Lisenfeld}},
		\bibinfo {author} {\bibfnamefont {C.}~\bibnamefont {Müller}}, \bibinfo
		{author} {\bibfnamefont {A.}~\bibnamefont {Bilmes}}, \bibinfo {author}
		{\bibfnamefont {A.}~\bibnamefont {Schneider}}, \bibinfo {author}
		{\bibfnamefont {D.~P.}\ \bibnamefont {Pappas}}, \bibinfo {author}
		{\bibfnamefont {A.~V.}\ \bibnamefont {Ustinov}},\ and\ \bibinfo {author}
		{\bibfnamefont {M.}~\bibnamefont {Weides}},\ }\bibfield  {title} {\bibinfo
		{title} {Correlating decoherence in transmon qubits: Low frequency noise by
			single fluctuators},\ }\href {https://doi.org/10.1103/PhysRevLett.123.190502}
	{\bibfield  {journal} {\bibinfo  {journal} {Phys. Rev. Lett.}\ }\textbf
		{\bibinfo {volume} {123}},\ \bibinfo {pages} {190502} (\bibinfo {year}
		{2019})}\BibitemShut {NoStop}%
	\bibitem [{\citenamefont {Burnett}\ \emph {et~al.}(2019)\citenamefont
		{Burnett}, \citenamefont {Bengtsson}, \citenamefont {Scigliuzzo},
		\citenamefont {Niepce}, \citenamefont {Kudra}, \citenamefont {Delsing},\ and\
		\citenamefont {Bylander}}]{burnett_decoherence_2019}%
	\BibitemOpen
	\bibfield  {author} {\bibinfo {author} {\bibfnamefont {J.~J.}\ \bibnamefont
			{Burnett}}, \bibinfo {author} {\bibfnamefont {A.}~\bibnamefont {Bengtsson}},
		\bibinfo {author} {\bibfnamefont {M.}~\bibnamefont {Scigliuzzo}}, \bibinfo
		{author} {\bibfnamefont {D.}~\bibnamefont {Niepce}}, \bibinfo {author}
		{\bibfnamefont {M.}~\bibnamefont {Kudra}}, \bibinfo {author} {\bibfnamefont
			{P.}~\bibnamefont {Delsing}},\ and\ \bibinfo {author} {\bibfnamefont
			{J.}~\bibnamefont {Bylander}},\ }\bibfield  {title} {\bibinfo {title}
		{Decoherence benchmarking of superconducting qubits},\ }\href@noop {}
	{\bibfield  {journal} {\bibinfo  {journal} {npj Quantum Inf.}\ }\textbf
		{\bibinfo {volume} {5}} (\bibinfo {year} {2019})}\BibitemShut {NoStop}%
	\bibitem [{\citenamefont {Neill}\ \emph {et~al.}(2013)\citenamefont {Neill},
		\citenamefont {Megrant}, \citenamefont {Barends}, \citenamefont {Chen},
		\citenamefont {Chiaro}, \citenamefont {Kelly}, \citenamefont {Mutus},
		\citenamefont {O'Malley}, \citenamefont {Sank}, \citenamefont {Wenner} \emph
		{et~al.}}]{neill_2013_fluctuations}%
	\BibitemOpen
	\bibfield  {author} {\bibinfo {author} {\bibfnamefont {C.}~\bibnamefont
			{Neill}}, \bibinfo {author} {\bibfnamefont {A.}~\bibnamefont {Megrant}},
		\bibinfo {author} {\bibfnamefont {R.}~\bibnamefont {Barends}}, \bibinfo
		{author} {\bibfnamefont {Y.}~\bibnamefont {Chen}}, \bibinfo {author}
		{\bibfnamefont {B.}~\bibnamefont {Chiaro}}, \bibinfo {author} {\bibfnamefont
			{J.}~\bibnamefont {Kelly}}, \bibinfo {author} {\bibfnamefont
			{J.}~\bibnamefont {Mutus}}, \bibinfo {author} {\bibfnamefont
			{P.}~\bibnamefont {O'Malley}}, \bibinfo {author} {\bibfnamefont
			{D.}~\bibnamefont {Sank}}, \bibinfo {author} {\bibfnamefont {J.}~\bibnamefont
			{Wenner}}, \emph {et~al.},\ }\bibfield  {title} {\bibinfo {title}
		{Fluctuations from edge defects in superconducting resonators},\ }\href@noop
	{} {\bibfield  {journal} {\bibinfo  {journal} {App. Phys. Lett.}\ }\textbf
		{\bibinfo {volume} {103}},\ \bibinfo {pages} {072601} (\bibinfo {year}
		{2013})}\BibitemShut {NoStop}%
	\bibitem [{\citenamefont {B{\'e}janin}\ \emph {et~al.}(2022)\citenamefont
		{B{\'e}janin}, \citenamefont {Ayadi}, \citenamefont {Xu}, \citenamefont
		{Zhu}, \citenamefont {Mohebbi},\ and\ \citenamefont
		{Mariantoni}}]{bejanin_2022_fluctuation}%
	\BibitemOpen
	\bibfield  {author} {\bibinfo {author} {\bibfnamefont {J.}~\bibnamefont
			{B{\'e}janin}}, \bibinfo {author} {\bibfnamefont {Y.}~\bibnamefont {Ayadi}},
		\bibinfo {author} {\bibfnamefont {X.}~\bibnamefont {Xu}}, \bibinfo {author}
		{\bibfnamefont {C.}~\bibnamefont {Zhu}}, \bibinfo {author} {\bibfnamefont
			{H.}~\bibnamefont {Mohebbi}},\ and\ \bibinfo {author} {\bibfnamefont
			{M.}~\bibnamefont {Mariantoni}},\ }\bibfield  {title} {\bibinfo {title}
		{Fluctuation spectroscopy of two-level systems in superconducting
			resonators},\ }\href@noop {} {\bibfield  {journal} {\bibinfo  {journal}
			{arXiv:2203.07445}\ } (\bibinfo {year} {2022})}\BibitemShut {NoStop}%
	\bibitem [{\citenamefont {de~Graaf}\ \emph {et~al.}(2021)\citenamefont
		{de~Graaf}, \citenamefont {Mahashabde}, \citenamefont {Kubatkin},
		\citenamefont {Tzalenchuk},\ and\ \citenamefont
		{Danilov}}]{deGraaf_2021_quantifying}%
	\BibitemOpen
	\bibfield  {author} {\bibinfo {author} {\bibfnamefont {S.}~\bibnamefont
			{de~Graaf}}, \bibinfo {author} {\bibfnamefont {S.}~\bibnamefont
			{Mahashabde}}, \bibinfo {author} {\bibfnamefont {S.}~\bibnamefont
			{Kubatkin}}, \bibinfo {author} {\bibfnamefont {A.~Y.}\ \bibnamefont
			{Tzalenchuk}},\ and\ \bibinfo {author} {\bibfnamefont {A.}~\bibnamefont
			{Danilov}},\ }\bibfield  {title} {\bibinfo {title} {Quantifying dynamics and
			interactions of individual spurious low-energy fluctuators in superconducting
			circuits},\ }\href@noop {} {\bibfield  {journal} {\bibinfo  {journal} {Phys.
				Rev. B}\ }\textbf {\bibinfo {volume} {103}},\ \bibinfo {pages} {174103}
		(\bibinfo {year} {2021})}\BibitemShut {NoStop}%
	\bibitem [{\citenamefont {Lisenfeld}\ \emph {et~al.}(2015)\citenamefont
		{Lisenfeld}, \citenamefont {Grabovskij}, \citenamefont {M{\"u}ller},
		\citenamefont {Cole}, \citenamefont {Weiss},\ and\ \citenamefont
		{Ustinov}}]{lisenfeld_2015_observation}%
	\BibitemOpen
	\bibfield  {author} {\bibinfo {author} {\bibfnamefont {J.}~\bibnamefont
			{Lisenfeld}}, \bibinfo {author} {\bibfnamefont {G.~J.}\ \bibnamefont
			{Grabovskij}}, \bibinfo {author} {\bibfnamefont {C.}~\bibnamefont
			{M{\"u}ller}}, \bibinfo {author} {\bibfnamefont {J.~H.}\ \bibnamefont
			{Cole}}, \bibinfo {author} {\bibfnamefont {G.}~\bibnamefont {Weiss}},\ and\
		\bibinfo {author} {\bibfnamefont {A.~V.}\ \bibnamefont {Ustinov}},\
	}\bibfield  {title} {\bibinfo {title} {Observation of directly interacting
			coherent two-level systems in an amorphous material},\ }\href@noop {}
	{\bibfield  {journal} {\bibinfo  {journal} {Nat. Commun.}\ }\textbf {\bibinfo
			{volume} {6}},\ \bibinfo {pages} {1} (\bibinfo {year} {2015})}\BibitemShut
	{NoStop}%
	\bibitem [{\citenamefont {Vepsäläinen}\ \emph {et~al.}(2020)\citenamefont
		{Vepsäläinen}, \citenamefont {Karamlou}, \citenamefont {Orrell},
		\citenamefont {Dogra}, \citenamefont {Loer}, \citenamefont {Vasconcelos},
		\citenamefont {Kim}, \citenamefont {Melville}, \citenamefont {Niedzielski},
		\citenamefont {Yoder}, \citenamefont {Gustavsson}, \citenamefont {Formaggio},
		\citenamefont {VanDevender},\ and\ \citenamefont
		{Oliver}}]{vepsalainen_impact_2020}%
	\BibitemOpen
	\bibfield  {author} {\bibinfo {author} {\bibfnamefont {A.~P.}\ \bibnamefont
			{Vepsäläinen}}, \bibinfo {author} {\bibfnamefont {A.~H.}\ \bibnamefont
			{Karamlou}}, \bibinfo {author} {\bibfnamefont {J.~L.}\ \bibnamefont
			{Orrell}}, \bibinfo {author} {\bibfnamefont {A.~S.}\ \bibnamefont {Dogra}},
		\bibinfo {author} {\bibfnamefont {B.}~\bibnamefont {Loer}}, \bibinfo {author}
		{\bibfnamefont {F.}~\bibnamefont {Vasconcelos}}, \bibinfo {author}
		{\bibfnamefont {D.~K.}\ \bibnamefont {Kim}}, \bibinfo {author} {\bibfnamefont
			{A.~J.}\ \bibnamefont {Melville}}, \bibinfo {author} {\bibfnamefont {B.~M.}\
			\bibnamefont {Niedzielski}}, \bibinfo {author} {\bibfnamefont {J.~L.}\
			\bibnamefont {Yoder}}, \bibinfo {author} {\bibfnamefont {S.}~\bibnamefont
			{Gustavsson}}, \bibinfo {author} {\bibfnamefont {J.~A.}\ \bibnamefont
			{Formaggio}}, \bibinfo {author} {\bibfnamefont {B.~A.}\ \bibnamefont
			{VanDevender}},\ and\ \bibinfo {author} {\bibfnamefont {W.~D.}\ \bibnamefont
			{Oliver}},\ }\bibfield  {title} {\bibinfo {title} {Impact of ionizing
			radiation on superconducting qubit coherence},\ }\href@noop {} {\bibfield
		{journal} {\bibinfo  {journal} {Nature}\ }\textbf {\bibinfo {volume} {584}}
		(\bibinfo {year} {2020})}\BibitemShut {NoStop}%
	\bibitem [{\citenamefont {Grünhaupt}\ \emph {et~al.}(2018)\citenamefont
		{Grünhaupt}, \citenamefont {Maleeva}, \citenamefont {Skacel}, \citenamefont
		{Calvo}, \citenamefont {Levy-Bertrand}, \citenamefont {Ustinov},
		\citenamefont {Rotzinger}, \citenamefont {Monfardini}, \citenamefont
		{Catelani},\ and\ \citenamefont {Pop}}]{grunhaupt_loss_2018}%
	\BibitemOpen
	\bibfield  {author} {\bibinfo {author} {\bibfnamefont {L.}~\bibnamefont
			{Grünhaupt}}, \bibinfo {author} {\bibfnamefont {N.}~\bibnamefont {Maleeva}},
		\bibinfo {author} {\bibfnamefont {S.~T.}\ \bibnamefont {Skacel}}, \bibinfo
		{author} {\bibfnamefont {M.}~\bibnamefont {Calvo}}, \bibinfo {author}
		{\bibfnamefont {F.}~\bibnamefont {Levy-Bertrand}}, \bibinfo {author}
		{\bibfnamefont {A.~V.}\ \bibnamefont {Ustinov}}, \bibinfo {author}
		{\bibfnamefont {H.}~\bibnamefont {Rotzinger}}, \bibinfo {author}
		{\bibfnamefont {A.}~\bibnamefont {Monfardini}}, \bibinfo {author}
		{\bibfnamefont {G.}~\bibnamefont {Catelani}},\ and\ \bibinfo {author}
		{\bibfnamefont {I.~M.}\ \bibnamefont {Pop}},\ }\bibfield  {title} {\bibinfo
		{title} {Loss mechanisms and quasiparticle dynamics in superconducting
			microwave resonators made of thin-film granular aluminum},\ }\href
	{https://doi.org/10.1103/PhysRevLett.121.117001} {\bibfield  {journal}
		{\bibinfo  {journal} {Phys. Rev. Lett.}\ }\textbf {\bibinfo {volume} {121}},\
		\bibinfo {pages} {117001} (\bibinfo {year} {2018})}\BibitemShut {NoStop}%
	\bibitem [{\citenamefont {Cardani}\ \emph {et~al.}(2021)\citenamefont
		{Cardani}, \citenamefont {Valenti}, \citenamefont {Casali}, \citenamefont
		{Catelani}, \citenamefont {Charpentier}, \citenamefont {Clemenza},
		\citenamefont {Colantoni}, \citenamefont {Cruciani}, \citenamefont
		{D’Imperio}, \citenamefont {Gironi} \emph
		{et~al.}}]{cardani_reducing_2021}%
	\BibitemOpen
	\bibfield  {author} {\bibinfo {author} {\bibfnamefont {L.}~\bibnamefont
			{Cardani}}, \bibinfo {author} {\bibfnamefont {F.}~\bibnamefont {Valenti}},
		\bibinfo {author} {\bibfnamefont {N.}~\bibnamefont {Casali}}, \bibinfo
		{author} {\bibfnamefont {G.}~\bibnamefont {Catelani}}, \bibinfo {author}
		{\bibfnamefont {T.}~\bibnamefont {Charpentier}}, \bibinfo {author}
		{\bibfnamefont {M.}~\bibnamefont {Clemenza}}, \bibinfo {author}
		{\bibfnamefont {I.}~\bibnamefont {Colantoni}}, \bibinfo {author}
		{\bibfnamefont {A.}~\bibnamefont {Cruciani}}, \bibinfo {author}
		{\bibfnamefont {G.}~\bibnamefont {D’Imperio}}, \bibinfo {author}
		{\bibfnamefont {L.}~\bibnamefont {Gironi}}, \emph {et~al.},\ }\bibfield
	{title} {\bibinfo {title} {Reducing the impact of radioactivity on quantum
			circuits in a deep-underground facility},\ }\href@noop {} {\bibfield
		{journal} {\bibinfo  {journal} {Nat. Commun.}\ }\textbf {\bibinfo {volume}
			{12}},\ \bibinfo {pages} {1} (\bibinfo {year} {2021})}\BibitemShut {NoStop}%
	\bibitem [{\citenamefont {Wilen}\ \emph {et~al.}(2021)\citenamefont {Wilen},
		\citenamefont {Abdullah}, \citenamefont {Kurinsky}, \citenamefont {Stanford},
		\citenamefont {Cardani}, \citenamefont {D’Imperio}, \citenamefont {Tomei},
		\citenamefont {Faoro}, \citenamefont {Ioffe}, \citenamefont {Liu},
		\citenamefont {Opremcak}, \citenamefont {Christensen}, \citenamefont
		{DuBois},\ and\ \citenamefont {McDermott}}]{wilen_correlated_2021}%
	\BibitemOpen
	\bibfield  {author} {\bibinfo {author} {\bibfnamefont {C.~D.}\ \bibnamefont
			{Wilen}}, \bibinfo {author} {\bibfnamefont {S.}~\bibnamefont {Abdullah}},
		\bibinfo {author} {\bibfnamefont {N.~A.}\ \bibnamefont {Kurinsky}}, \bibinfo
		{author} {\bibfnamefont {C.}~\bibnamefont {Stanford}}, \bibinfo {author}
		{\bibfnamefont {L.}~\bibnamefont {Cardani}}, \bibinfo {author} {\bibfnamefont
			{G.}~\bibnamefont {D’Imperio}}, \bibinfo {author} {\bibfnamefont
			{C.}~\bibnamefont {Tomei}}, \bibinfo {author} {\bibfnamefont
			{L.}~\bibnamefont {Faoro}}, \bibinfo {author} {\bibfnamefont {L.~B.}\
			\bibnamefont {Ioffe}}, \bibinfo {author} {\bibfnamefont {C.~H.}\ \bibnamefont
			{Liu}}, \bibinfo {author} {\bibfnamefont {A.}~\bibnamefont {Opremcak}},
		\bibinfo {author} {\bibfnamefont {B.~G.}\ \bibnamefont {Christensen}},
		\bibinfo {author} {\bibfnamefont {J.~L.}\ \bibnamefont {DuBois}},\ and\
		\bibinfo {author} {\bibfnamefont {R.}~\bibnamefont {McDermott}},\ }\bibfield
	{title} {\bibinfo {title} {Correlated charge noise and relaxation errors in
			superconducting qubits},\ }\href@noop {} {\bibfield  {journal} {\bibinfo
			{journal} {Nature}\ }\textbf {\bibinfo {volume} {594}},\ \bibinfo {pages}
		{369} (\bibinfo {year} {2021})}\BibitemShut {NoStop}%
	\bibitem [{\citenamefont {McEwen}\ \emph {et~al.}(2022)\citenamefont {McEwen},
		\citenamefont {Faoro}, \citenamefont {Arya}, \citenamefont {Dunsworth},
		\citenamefont {Huang}, \citenamefont {Kim}, \citenamefont {Burkett},
		\citenamefont {Fowler}, \citenamefont {Arute}, \citenamefont {Bardin} \emph
		{et~al.}}]{mcewen_resolving_2021}%
	\BibitemOpen
	\bibfield  {author} {\bibinfo {author} {\bibfnamefont {M.}~\bibnamefont
			{McEwen}}, \bibinfo {author} {\bibfnamefont {L.}~\bibnamefont {Faoro}},
		\bibinfo {author} {\bibfnamefont {K.}~\bibnamefont {Arya}}, \bibinfo {author}
		{\bibfnamefont {A.}~\bibnamefont {Dunsworth}}, \bibinfo {author}
		{\bibfnamefont {T.}~\bibnamefont {Huang}}, \bibinfo {author} {\bibfnamefont
			{S.}~\bibnamefont {Kim}}, \bibinfo {author} {\bibfnamefont {B.}~\bibnamefont
			{Burkett}}, \bibinfo {author} {\bibfnamefont {A.}~\bibnamefont {Fowler}},
		\bibinfo {author} {\bibfnamefont {F.}~\bibnamefont {Arute}}, \bibinfo
		{author} {\bibfnamefont {J.~C.}\ \bibnamefont {Bardin}}, \emph {et~al.},\
	}\bibfield  {title} {\bibinfo {title} {Resolving catastrophic error bursts
			from cosmic rays in large arrays of superconducting qubits},\ }\href@noop {}
	{\bibfield  {journal} {\bibinfo  {journal} {Nat. Phys.}\ }\textbf {\bibinfo
			{volume} {18}},\ \bibinfo {pages} {107} (\bibinfo {year} {2022})}\BibitemShut
	{NoStop}%
	\bibitem [{\citenamefont {Martinis}(2021)}]{martinis_saving_2021}%
	\BibitemOpen
	\bibfield  {author} {\bibinfo {author} {\bibfnamefont {J.~M.}\ \bibnamefont
			{Martinis}},\ }\bibfield  {title} {\bibinfo {title} {Saving superconducting
			quantum processors from decay and correlated errors generated by gamma and
			cosmic rays},\ }\href@noop {} {\bibfield  {journal} {\bibinfo  {journal} {npj
				Quantum Inf.}\ }\textbf {\bibinfo {volume} {7}},\ \bibinfo {pages} {1}
		(\bibinfo {year} {2021})}\BibitemShut {NoStop}%
	\bibitem [{\citenamefont {Aharonov}\ \emph {et~al.}(2006)\citenamefont
		{Aharonov}, \citenamefont {Kitaev},\ and\ \citenamefont
		{Preskill}}]{aharonov_2006_fault}%
	\BibitemOpen
	\bibfield  {author} {\bibinfo {author} {\bibfnamefont {D.}~\bibnamefont
			{Aharonov}}, \bibinfo {author} {\bibfnamefont {A.}~\bibnamefont {Kitaev}},\
		and\ \bibinfo {author} {\bibfnamefont {J.}~\bibnamefont {Preskill}},\
	}\bibfield  {title} {\bibinfo {title} {Fault-tolerant quantum computation
			with long-range correlated noise},\ }\href@noop {} {\bibfield  {journal}
		{\bibinfo  {journal} {Phys. Rev. Lett.}\ }\textbf {\bibinfo {volume} {96}},\
		\bibinfo {pages} {050504} (\bibinfo {year} {2006})}\BibitemShut {NoStop}%
	\bibitem [{\citenamefont {Chen}\ \emph {et~al.}(2021)\citenamefont {Chen},
		\citenamefont {Satzinger}, \citenamefont {Atalaya}, \citenamefont {Korotkov},
		\citenamefont {Dunsworth}, \citenamefont {Sank}, \citenamefont {Quintana},
		\citenamefont {McEwen}, \citenamefont {Barends}, \citenamefont {Klimov} \emph
		{et~al.}}]{chen2021exponential}%
	\BibitemOpen
	\bibfield  {author} {\bibinfo {author} {\bibfnamefont {Z.}~\bibnamefont
			{Chen}}, \bibinfo {author} {\bibfnamefont {K.~J.}\ \bibnamefont {Satzinger}},
		\bibinfo {author} {\bibfnamefont {J.}~\bibnamefont {Atalaya}}, \bibinfo
		{author} {\bibfnamefont {A.~N.}\ \bibnamefont {Korotkov}}, \bibinfo {author}
		{\bibfnamefont {A.}~\bibnamefont {Dunsworth}}, \bibinfo {author}
		{\bibfnamefont {D.}~\bibnamefont {Sank}}, \bibinfo {author} {\bibfnamefont
			{C.}~\bibnamefont {Quintana}}, \bibinfo {author} {\bibfnamefont
			{M.}~\bibnamefont {McEwen}}, \bibinfo {author} {\bibfnamefont
			{R.}~\bibnamefont {Barends}}, \bibinfo {author} {\bibfnamefont {P.~V.}\
			\bibnamefont {Klimov}}, \emph {et~al.},\ }\bibfield  {title} {\bibinfo
		{title} {Exponential suppression of bit or phase errors with cyclic error
			correction},\ }\href@noop {} {\bibfield  {journal} {\bibinfo  {journal}
			{Nature}\ }\textbf {\bibinfo {volume} {595}},\ \bibinfo {pages} {383}
		(\bibinfo {year} {2021})}\BibitemShut {NoStop}%
	\bibitem [{\citenamefont {Riwar}\ \emph {et~al.}(2016)\citenamefont {Riwar},
		\citenamefont {Hosseinkhani}, \citenamefont {Burkhart}, \citenamefont {Gao},
		\citenamefont {Schoelkopf}, \citenamefont {Glazman},\ and\ \citenamefont
		{Catelani}}]{riwar_2016_normal}%
	\BibitemOpen
	\bibfield  {author} {\bibinfo {author} {\bibfnamefont {R.-P.}\ \bibnamefont
			{Riwar}}, \bibinfo {author} {\bibfnamefont {A.}~\bibnamefont {Hosseinkhani}},
		\bibinfo {author} {\bibfnamefont {L.~D.}\ \bibnamefont {Burkhart}}, \bibinfo
		{author} {\bibfnamefont {Y.~Y.}\ \bibnamefont {Gao}}, \bibinfo {author}
		{\bibfnamefont {R.~J.}\ \bibnamefont {Schoelkopf}}, \bibinfo {author}
		{\bibfnamefont {L.~I.}\ \bibnamefont {Glazman}},\ and\ \bibinfo {author}
		{\bibfnamefont {G.}~\bibnamefont {Catelani}},\ }\bibfield  {title} {\bibinfo
		{title} {Normal-metal quasiparticle traps for superconducting qubits},\
	}\href@noop {} {\bibfield  {journal} {\bibinfo  {journal} {Phys. Rev. B}\
		}\textbf {\bibinfo {volume} {94}},\ \bibinfo {pages} {104516} (\bibinfo
		{year} {2016})}\BibitemShut {NoStop}%
	\bibitem [{\citenamefont {Hosseinkhani}\ and\ \citenamefont
		{Catelani}(2018)}]{hosseinkhani_2018_proximity}%
	\BibitemOpen
	\bibfield  {author} {\bibinfo {author} {\bibfnamefont {A.}~\bibnamefont
			{Hosseinkhani}}\ and\ \bibinfo {author} {\bibfnamefont {G.}~\bibnamefont
			{Catelani}},\ }\bibfield  {title} {\bibinfo {title} {Proximity effect in
			normal-metal quasiparticle traps},\ }\href@noop {} {\bibfield  {journal}
		{\bibinfo  {journal} {Phys. Rev. B}\ }\textbf {\bibinfo {volume} {97}},\
		\bibinfo {pages} {054513} (\bibinfo {year} {2018})}\BibitemShut {NoStop}%
	\bibitem [{\citenamefont {Patel}\ \emph {et~al.}(2017)\citenamefont {Patel},
		\citenamefont {Pechenezhskiy}, \citenamefont {Plourde}, \citenamefont
		{Vavilov},\ and\ \citenamefont {McDermott}}]{patel2017phonon}%
	\BibitemOpen
	\bibfield  {author} {\bibinfo {author} {\bibfnamefont {U.}~\bibnamefont
			{Patel}}, \bibinfo {author} {\bibfnamefont {I.~V.}\ \bibnamefont
			{Pechenezhskiy}}, \bibinfo {author} {\bibfnamefont {B.}~\bibnamefont
			{Plourde}}, \bibinfo {author} {\bibfnamefont {M.}~\bibnamefont {Vavilov}},\
		and\ \bibinfo {author} {\bibfnamefont {R.}~\bibnamefont {McDermott}},\
	}\bibfield  {title} {\bibinfo {title} {Phonon-mediated quasiparticle
			poisoning of superconducting microwave resonators},\ }\href@noop {}
	{\bibfield  {journal} {\bibinfo  {journal} {Phys. Rev. B}\ }\textbf {\bibinfo
			{volume} {96}},\ \bibinfo {pages} {220501} (\bibinfo {year}
		{2017})}\BibitemShut {NoStop}%
	\bibitem [{\citenamefont {Court}\ \emph {et~al.}(2008)\citenamefont {Court},
		\citenamefont {Ferguson}, \citenamefont {Lutchyn},\ and\ \citenamefont
		{Clark}}]{court2008quantitative}%
	\BibitemOpen
	\bibfield  {author} {\bibinfo {author} {\bibfnamefont {N.}~\bibnamefont
			{Court}}, \bibinfo {author} {\bibfnamefont {A.}~\bibnamefont {Ferguson}},
		\bibinfo {author} {\bibfnamefont {R.}~\bibnamefont {Lutchyn}},\ and\ \bibinfo
		{author} {\bibfnamefont {R.}~\bibnamefont {Clark}},\ }\bibfield  {title}
	{\bibinfo {title} {Quantitative study of quasiparticle traps using the
			single-cooper-pair transistor},\ }\href@noop {} {\bibfield  {journal}
		{\bibinfo  {journal} {Phys. Rev. B}\ }\textbf {\bibinfo {volume} {77}},\
		\bibinfo {pages} {100501} (\bibinfo {year} {2008})}\BibitemShut {NoStop}%
	\bibitem [{\citenamefont {Iaia}\ \emph {et~al.}(2022)\citenamefont {Iaia},
		\citenamefont {Ku}, \citenamefont {Ballard}, \citenamefont {Larson},
		\citenamefont {Yelton}, \citenamefont {Liu}, \citenamefont {Patel},
		\citenamefont {McDermott},\ and\ \citenamefont {Plourde}}]{iaia_2022_phonon}%
	\BibitemOpen
	\bibfield  {author} {\bibinfo {author} {\bibfnamefont {V.}~\bibnamefont
			{Iaia}}, \bibinfo {author} {\bibfnamefont {J.}~\bibnamefont {Ku}}, \bibinfo
		{author} {\bibfnamefont {A.}~\bibnamefont {Ballard}}, \bibinfo {author}
		{\bibfnamefont {C.}~\bibnamefont {Larson}}, \bibinfo {author} {\bibfnamefont
			{E.}~\bibnamefont {Yelton}}, \bibinfo {author} {\bibfnamefont
			{C.}~\bibnamefont {Liu}}, \bibinfo {author} {\bibfnamefont {S.}~\bibnamefont
			{Patel}}, \bibinfo {author} {\bibfnamefont {R.}~\bibnamefont {McDermott}},\
		and\ \bibinfo {author} {\bibfnamefont {B.}~\bibnamefont {Plourde}},\
	}\bibfield  {title} {\bibinfo {title} {Phonon downconversion to suppress
			correlated errors in superconducting qubits},\ }\href@noop {} {\bibfield
		{journal} {\bibinfo  {journal} {arXiv:2203.06586}\ } (\bibinfo {year}
		{2022})}\BibitemShut {NoStop}%
	\bibitem [{\citenamefont {Henriques}\ \emph {et~al.}(2019)\citenamefont
		{Henriques}, \citenamefont {Valenti}, \citenamefont {Charpentier},
		\citenamefont {Lagoin}, \citenamefont {Gouriou}, \citenamefont
		{Mart{\'\i}nez}, \citenamefont {Cardani}, \citenamefont {Vignati},
		\citenamefont {Gr{\"u}nhaupt}, \citenamefont {Gusenkova} \emph
		{et~al.}}]{henriques_2019_phonon}%
	\BibitemOpen
	\bibfield  {author} {\bibinfo {author} {\bibfnamefont {F.}~\bibnamefont
			{Henriques}}, \bibinfo {author} {\bibfnamefont {F.}~\bibnamefont {Valenti}},
		\bibinfo {author} {\bibfnamefont {T.}~\bibnamefont {Charpentier}}, \bibinfo
		{author} {\bibfnamefont {M.}~\bibnamefont {Lagoin}}, \bibinfo {author}
		{\bibfnamefont {C.}~\bibnamefont {Gouriou}}, \bibinfo {author} {\bibfnamefont
			{M.}~\bibnamefont {Mart{\'\i}nez}}, \bibinfo {author} {\bibfnamefont
			{L.}~\bibnamefont {Cardani}}, \bibinfo {author} {\bibfnamefont
			{M.}~\bibnamefont {Vignati}}, \bibinfo {author} {\bibfnamefont
			{L.}~\bibnamefont {Gr{\"u}nhaupt}}, \bibinfo {author} {\bibfnamefont
			{D.}~\bibnamefont {Gusenkova}}, \emph {et~al.},\ }\bibfield  {title}
	{\bibinfo {title} {Phonon traps reduce the quasiparticle density in
			superconducting circuits},\ }\href@noop {} {\bibfield  {journal} {\bibinfo
			{journal} {App. Phys. Lett.}\ }\textbf {\bibinfo {volume} {115}},\ \bibinfo
		{pages} {212601} (\bibinfo {year} {2019})}\BibitemShut {NoStop}%
	\bibitem [{\citenamefont {Orrell}\ and\ \citenamefont
		{Loer}(2021)}]{orrell_sensor-assisted_2021}%
	\BibitemOpen
	\bibfield  {author} {\bibinfo {author} {\bibfnamefont {J.~L.}\ \bibnamefont
			{Orrell}}\ and\ \bibinfo {author} {\bibfnamefont {B.}~\bibnamefont {Loer}},\
	}\bibfield  {title} {\bibinfo {title} {Sensor-assisted fault mitigation in
			quantum computation},\ }\href@noop {} {\bibfield  {journal} {\bibinfo
			{journal} {Phys. Rev. App.}\ }\textbf {\bibinfo {volume} {16}},\ \bibinfo
		{pages} {024025} (\bibinfo {year} {2021})}\BibitemShut {NoStop}%
	\bibitem [{\citenamefont {Xu}\ \emph {et~al.}(2022)\citenamefont {Xu},
		\citenamefont {Seif}, \citenamefont {Yan}, \citenamefont {Mannucci},
		\citenamefont {Sane}, \citenamefont {Van~Meter}, \citenamefont {Cleland},\
		and\ \citenamefont {Jiang}}]{xu_2022_distributed}%
	\BibitemOpen
	\bibfield  {author} {\bibinfo {author} {\bibfnamefont {Q.}~\bibnamefont
			{Xu}}, \bibinfo {author} {\bibfnamefont {A.}~\bibnamefont {Seif}}, \bibinfo
		{author} {\bibfnamefont {H.}~\bibnamefont {Yan}}, \bibinfo {author}
		{\bibfnamefont {N.}~\bibnamefont {Mannucci}}, \bibinfo {author}
		{\bibfnamefont {B.~O.}\ \bibnamefont {Sane}}, \bibinfo {author}
		{\bibfnamefont {R.}~\bibnamefont {Van~Meter}}, \bibinfo {author}
		{\bibfnamefont {A.~N.}\ \bibnamefont {Cleland}},\ and\ \bibinfo {author}
		{\bibfnamefont {L.}~\bibnamefont {Jiang}},\ }\bibfield  {title} {\bibinfo
		{title} {Distributed quantum error correction for chip-level catastrophic
			errors},\ }\href@noop {} {\bibfield  {journal} {\bibinfo  {journal}
			{arXiv:2203.16488}\ } (\bibinfo {year} {2022})}\BibitemShut {NoStop}%
	\bibitem [{\citenamefont {Leman}(2012)}]{leman_2012_invited}%
	\BibitemOpen
	\bibfield  {author} {\bibinfo {author} {\bibfnamefont {S.~W.}\ \bibnamefont
			{Leman}},\ }\bibfield  {title} {\bibinfo {title} {Physics and monte carlo
			techniques as relevant to cryogenic, phonon, and ionization readout of
			cryogenic dark matter search radiation detectors},\ }\href@noop {} {\bibfield
		{journal} {\bibinfo  {journal} {Rev. Sci. Instrum.}\ }\textbf {\bibinfo
			{volume} {83}},\ \bibinfo {pages} {091101} (\bibinfo {year}
		{2012})}\BibitemShut {NoStop}%
	\bibitem [{\citenamefont {Gordon}\ \emph {et~al.}(2022)\citenamefont {Gordon},
		\citenamefont {Murray}, \citenamefont {Kurter}, \citenamefont {Sandberg},
		\citenamefont {Hall}, \citenamefont {Balakrishnan}, \citenamefont {Shelby},
		\citenamefont {Wacaser}, \citenamefont {Stabile}, \citenamefont {Sleight}
		\emph {et~al.}}]{gordon_2022_environmental}%
	\BibitemOpen
	\bibfield  {author} {\bibinfo {author} {\bibfnamefont {R.}~\bibnamefont
			{Gordon}}, \bibinfo {author} {\bibfnamefont {C.}~\bibnamefont {Murray}},
		\bibinfo {author} {\bibfnamefont {C.}~\bibnamefont {Kurter}}, \bibinfo
		{author} {\bibfnamefont {M.}~\bibnamefont {Sandberg}}, \bibinfo {author}
		{\bibfnamefont {S.}~\bibnamefont {Hall}}, \bibinfo {author} {\bibfnamefont
			{K.}~\bibnamefont {Balakrishnan}}, \bibinfo {author} {\bibfnamefont
			{R.}~\bibnamefont {Shelby}}, \bibinfo {author} {\bibfnamefont
			{B.}~\bibnamefont {Wacaser}}, \bibinfo {author} {\bibfnamefont
			{A.}~\bibnamefont {Stabile}}, \bibinfo {author} {\bibfnamefont
			{J.}~\bibnamefont {Sleight}}, \emph {et~al.},\ }\bibfield  {title} {\bibinfo
		{title} {Environmental radiation impact on lifetimes and quasiparticle
			tunneling rates of fixed-frequency transmon qubits},\ }\href@noop {}
	{\bibfield  {journal} {\bibinfo  {journal} {App. Phys. Lett.}\ }\textbf
		{\bibinfo {volume} {120}},\ \bibinfo {pages} {074002} (\bibinfo {year}
		{2022})}\BibitemShut {NoStop}%
	\bibitem [{\citenamefont {C{\'o}rcoles}\ \emph {et~al.}(2011)\citenamefont
		{C{\'o}rcoles}, \citenamefont {Chow}, \citenamefont {Gambetta}, \citenamefont
		{Rigetti}, \citenamefont {Rozen}, \citenamefont {Keefe}, \citenamefont
		{Beth~Rothwell}, \citenamefont {Ketchen},\ and\ \citenamefont
		{Steffen}}]{corcoles_2011_protecting}%
	\BibitemOpen
	\bibfield  {author} {\bibinfo {author} {\bibfnamefont {A.~D.}\ \bibnamefont
			{C{\'o}rcoles}}, \bibinfo {author} {\bibfnamefont {J.~M.}\ \bibnamefont
			{Chow}}, \bibinfo {author} {\bibfnamefont {J.~M.}\ \bibnamefont {Gambetta}},
		\bibinfo {author} {\bibfnamefont {C.}~\bibnamefont {Rigetti}}, \bibinfo
		{author} {\bibfnamefont {J.~R.}\ \bibnamefont {Rozen}}, \bibinfo {author}
		{\bibfnamefont {G.~A.}\ \bibnamefont {Keefe}}, \bibinfo {author}
		{\bibfnamefont {M.}~\bibnamefont {Beth~Rothwell}}, \bibinfo {author}
		{\bibfnamefont {M.~B.}\ \bibnamefont {Ketchen}},\ and\ \bibinfo {author}
		{\bibfnamefont {M.}~\bibnamefont {Steffen}},\ }\bibfield  {title} {\bibinfo
		{title} {Protecting superconducting qubits from radiation},\ }\href@noop {}
	{\bibfield  {journal} {\bibinfo  {journal} {App. Phys. Lett.}\ }\textbf
		{\bibinfo {volume} {99}},\ \bibinfo {pages} {181906} (\bibinfo {year}
		{2011})}\BibitemShut {NoStop}%
	\bibitem [{\citenamefont {Barends}\ \emph {et~al.}(2011)\citenamefont
		{Barends}, \citenamefont {Wenner}, \citenamefont {Lenander}, \citenamefont
		{Chen}, \citenamefont {Bialczak}, \citenamefont {Kelly}, \citenamefont
		{Lucero}, \citenamefont {O’Malley}, \citenamefont {Mariantoni},
		\citenamefont {Sank} \emph {et~al.}}]{barends_2011_minimizing}%
	\BibitemOpen
	\bibfield  {author} {\bibinfo {author} {\bibfnamefont {R.}~\bibnamefont
			{Barends}}, \bibinfo {author} {\bibfnamefont {J.}~\bibnamefont {Wenner}},
		\bibinfo {author} {\bibfnamefont {M.}~\bibnamefont {Lenander}}, \bibinfo
		{author} {\bibfnamefont {Y.}~\bibnamefont {Chen}}, \bibinfo {author}
		{\bibfnamefont {R.~C.}\ \bibnamefont {Bialczak}}, \bibinfo {author}
		{\bibfnamefont {J.}~\bibnamefont {Kelly}}, \bibinfo {author} {\bibfnamefont
			{E.}~\bibnamefont {Lucero}}, \bibinfo {author} {\bibfnamefont
			{P.}~\bibnamefont {O’Malley}}, \bibinfo {author} {\bibfnamefont
			{M.}~\bibnamefont {Mariantoni}}, \bibinfo {author} {\bibfnamefont
			{D.}~\bibnamefont {Sank}}, \emph {et~al.},\ }\bibfield  {title} {\bibinfo
		{title} {Minimizing quasiparticle generation from stray infrared light in
			superconducting quantum circuits},\ }\href@noop {} {\bibfield  {journal}
		{\bibinfo  {journal} {App. Phys. Lett.}\ }\textbf {\bibinfo {volume} {99}},\
		\bibinfo {pages} {113507} (\bibinfo {year} {2011})}\BibitemShut {NoStop}%
	\bibitem [{\citenamefont {Baselmans}(2012)}]{baselmans_kinetic_2012}%
	\BibitemOpen
	\bibfield  {author} {\bibinfo {author} {\bibfnamefont {J.}~\bibnamefont
			{Baselmans}},\ }\bibfield  {title} {\bibinfo {title} {Kinetic {Inductance}
			{Detectors}},\ }\href@noop {} {\bibfield  {journal} {\bibinfo  {journal} {J.
				of Low Temp. Phys.}\ }\textbf {\bibinfo {volume} {167}},\ \bibinfo {pages}
		{292} (\bibinfo {year} {2012})}\BibitemShut {NoStop}%
	\bibitem [{\citenamefont {Tennant}\ \emph {et~al.}(2022)\citenamefont
		{Tennant}, \citenamefont {Martinez}, \citenamefont {Beck}, \citenamefont
		{O'Kelley}, \citenamefont {Wilen}, \citenamefont {McDermott}, \citenamefont
		{DuBois},\ and\ \citenamefont {Rosen}}]{tennant_low_2021}%
	\BibitemOpen
	\bibfield  {author} {\bibinfo {author} {\bibfnamefont {D.~M.}\ \bibnamefont
			{Tennant}}, \bibinfo {author} {\bibfnamefont {L.~A.}\ \bibnamefont
			{Martinez}}, \bibinfo {author} {\bibfnamefont {K.~M.}\ \bibnamefont {Beck}},
		\bibinfo {author} {\bibfnamefont {S.~R.}\ \bibnamefont {O'Kelley}}, \bibinfo
		{author} {\bibfnamefont {C.~D.}\ \bibnamefont {Wilen}}, \bibinfo {author}
		{\bibfnamefont {R.}~\bibnamefont {McDermott}}, \bibinfo {author}
		{\bibfnamefont {J.~L.}\ \bibnamefont {DuBois}},\ and\ \bibinfo {author}
		{\bibfnamefont {Y.~J.}\ \bibnamefont {Rosen}},\ }\bibfield  {title} {\bibinfo
		{title} {Low-frequency correlated charge-noise measurements across multiple
			energy transitions in a tantalum transmon},\ }\href@noop {} {\bibfield
		{journal} {\bibinfo  {journal} {PRX Quantum}\ }\textbf {\bibinfo {volume}
			{3}},\ \bibinfo {pages} {030307} (\bibinfo {year} {2022})}\BibitemShut
	{NoStop}%
	\bibitem [{\citenamefont {Ristè}\ \emph {et~al.}(2013)\citenamefont {Ristè},
		\citenamefont {Bultink}, \citenamefont {Tiggelman}, \citenamefont {Schouten},
		\citenamefont {Lehnert},\ and\ \citenamefont
		{DiCarlo}}]{riste_millisecond_2013}%
	\BibitemOpen
	\bibfield  {author} {\bibinfo {author} {\bibfnamefont {D.}~\bibnamefont
			{Ristè}}, \bibinfo {author} {\bibfnamefont {C.~C.}\ \bibnamefont {Bultink}},
		\bibinfo {author} {\bibfnamefont {M.~J.}\ \bibnamefont {Tiggelman}}, \bibinfo
		{author} {\bibfnamefont {R.~N.}\ \bibnamefont {Schouten}}, \bibinfo {author}
		{\bibfnamefont {K.~W.}\ \bibnamefont {Lehnert}},\ and\ \bibinfo {author}
		{\bibfnamefont {L.}~\bibnamefont {DiCarlo}},\ }\bibfield  {title} {\bibinfo
		{title} {Millisecond charge-parity fluctuations and induced decoherence in a
			superconducting transmon qubit},\ }\href {https://doi.org/10.1038/ncomms2936}
	{\bibfield  {journal} {\bibinfo  {journal} {Nat. Commun.}\ }\textbf {\bibinfo
			{volume} {4}},\ \bibinfo {pages} {1913} (\bibinfo {year} {2013})}\BibitemShut
	{NoStop}%
	\bibitem [{\citenamefont {Stewart}\ and\ \citenamefont
		{Zimmerman}(2016)}]{stewart_2016_stability}%
	\BibitemOpen
	\bibfield  {author} {\bibinfo {author} {\bibfnamefont {M.~D.}\ \bibnamefont
			{Stewart}}\ and\ \bibinfo {author} {\bibfnamefont {N.~M.}\ \bibnamefont
			{Zimmerman}},\ }\bibfield  {title} {\bibinfo {title} {Stability of single
			electron devices: charge offset drift},\ }\href@noop {} {\bibfield  {journal}
		{\bibinfo  {journal} {App. Sci.}\ }\textbf {\bibinfo {volume} {6}},\ \bibinfo
		{pages} {187} (\bibinfo {year} {2016})}\BibitemShut {NoStop}%
	\bibitem [{\citenamefont {Zimmerman}\ \emph {et~al.}(2008)\citenamefont
		{Zimmerman}, \citenamefont {Huber}, \citenamefont {Simonds}, \citenamefont
		{Hourdakis}, \citenamefont {Fujiwara}, \citenamefont {Ono}, \citenamefont
		{Takahashi}, \citenamefont {Inokawa}, \citenamefont {Furlan},\ and\
		\citenamefont {Keller}}]{zimmerman2008long}%
	\BibitemOpen
	\bibfield  {author} {\bibinfo {author} {\bibfnamefont {N.~M.}\ \bibnamefont
			{Zimmerman}}, \bibinfo {author} {\bibfnamefont {W.~H.}\ \bibnamefont
			{Huber}}, \bibinfo {author} {\bibfnamefont {B.}~\bibnamefont {Simonds}},
		\bibinfo {author} {\bibfnamefont {E.}~\bibnamefont {Hourdakis}}, \bibinfo
		{author} {\bibfnamefont {A.}~\bibnamefont {Fujiwara}}, \bibinfo {author}
		{\bibfnamefont {Y.}~\bibnamefont {Ono}}, \bibinfo {author} {\bibfnamefont
			{Y.}~\bibnamefont {Takahashi}}, \bibinfo {author} {\bibfnamefont
			{H.}~\bibnamefont {Inokawa}}, \bibinfo {author} {\bibfnamefont
			{M.}~\bibnamefont {Furlan}},\ and\ \bibinfo {author} {\bibfnamefont {M.~W.}\
			\bibnamefont {Keller}},\ }\bibfield  {title} {\bibinfo {title} {Why the
			long-term charge offset drift in si single-electron tunneling transistors is
			much smaller (better) than in metal-based ones: Two-level fluctuator
			stability},\ }\href@noop {} {\bibfield  {journal} {\bibinfo  {journal} {J.
				App. Phys.}\ }\textbf {\bibinfo {volume} {104}},\ \bibinfo {pages} {033710}
		(\bibinfo {year} {2008})}\BibitemShut {NoStop}%
	\bibitem [{\citenamefont {Kafanov}\ \emph {et~al.}(2008)\citenamefont
		{Kafanov}, \citenamefont {Brenning}, \citenamefont {Duty},\ and\
		\citenamefont {Delsing}}]{kafanov2008charge}%
	\BibitemOpen
	\bibfield  {author} {\bibinfo {author} {\bibfnamefont {S.}~\bibnamefont
			{Kafanov}}, \bibinfo {author} {\bibfnamefont {H.}~\bibnamefont {Brenning}},
		\bibinfo {author} {\bibfnamefont {T.}~\bibnamefont {Duty}},\ and\ \bibinfo
		{author} {\bibfnamefont {P.}~\bibnamefont {Delsing}},\ }\bibfield  {title}
	{\bibinfo {title} {Charge noise in single-electron transistors and charge
			qubits may be caused by metallic grains},\ }\href@noop {} {\bibfield
		{journal} {\bibinfo  {journal} {Phys. Rev. B}\ }\textbf {\bibinfo {volume}
			{78}},\ \bibinfo {pages} {125411} (\bibinfo {year} {2008})}\BibitemShut
	{NoStop}%
	\bibitem [{\citenamefont {Christensen}\ \emph {et~al.}(2019)\citenamefont
		{Christensen}, \citenamefont {Wilen}, \citenamefont {Opremcak}, \citenamefont
		{Nelson}, \citenamefont {Schlenker}, \citenamefont {Zimonick}, \citenamefont
		{Faoro}, \citenamefont {Ioffe}, \citenamefont {Rosen}, \citenamefont
		{DuBois}, \citenamefont {Plourde},\ and\ \citenamefont
		{McDermott}}]{christensen_anomalous_2019}%
	\BibitemOpen
	\bibfield  {author} {\bibinfo {author} {\bibfnamefont {B.~G.}\ \bibnamefont
			{Christensen}}, \bibinfo {author} {\bibfnamefont {C.~D.}\ \bibnamefont
			{Wilen}}, \bibinfo {author} {\bibfnamefont {A.}~\bibnamefont {Opremcak}},
		\bibinfo {author} {\bibfnamefont {J.}~\bibnamefont {Nelson}}, \bibinfo
		{author} {\bibfnamefont {F.}~\bibnamefont {Schlenker}}, \bibinfo {author}
		{\bibfnamefont {C.~H.}\ \bibnamefont {Zimonick}}, \bibinfo {author}
		{\bibfnamefont {L.}~\bibnamefont {Faoro}}, \bibinfo {author} {\bibfnamefont
			{L.~B.}\ \bibnamefont {Ioffe}}, \bibinfo {author} {\bibfnamefont {Y.~J.}\
			\bibnamefont {Rosen}}, \bibinfo {author} {\bibfnamefont {J.~L.}\ \bibnamefont
			{DuBois}}, \bibinfo {author} {\bibfnamefont {B.~L.~T.}\ \bibnamefont
			{Plourde}},\ and\ \bibinfo {author} {\bibfnamefont {R.}~\bibnamefont
			{McDermott}},\ }\bibfield  {title} {\bibinfo {title} {Anomalous charge noise
			in superconducting qubits},\ }\href
	{https://doi.org/10.1103/PhysRevB.100.140503} {\bibfield  {journal} {\bibinfo
			{journal} {Phys. Rev. B}\ }\textbf {\bibinfo {volume} {100}},\ \bibinfo
		{pages} {140503} (\bibinfo {year} {2019})}\BibitemShut {NoStop}%
	\bibitem [{\citenamefont {Pourkabirian}\ \emph {et~al.}(2014)\citenamefont
		{Pourkabirian}, \citenamefont {Gustafsson}, \citenamefont {Johansson},
		\citenamefont {Clarke},\ and\ \citenamefont
		{Delsing}}]{pourkabirian_2014_nonequilibrium}%
	\BibitemOpen
	\bibfield  {author} {\bibinfo {author} {\bibfnamefont {A.}~\bibnamefont
			{Pourkabirian}}, \bibinfo {author} {\bibfnamefont {M.~V.}\ \bibnamefont
			{Gustafsson}}, \bibinfo {author} {\bibfnamefont {G.}~\bibnamefont
			{Johansson}}, \bibinfo {author} {\bibfnamefont {J.}~\bibnamefont {Clarke}},\
		and\ \bibinfo {author} {\bibfnamefont {P.}~\bibnamefont {Delsing}},\
	}\bibfield  {title} {\bibinfo {title} {Nonequilibrium probing of two-level
			charge fluctuators using the step response of a single-electron transistor},\
	}\href@noop {} {\bibfield  {journal} {\bibinfo  {journal} {Phys. Rev. Lett.}\
		}\textbf {\bibinfo {volume} {113}},\ \bibinfo {pages} {256801} (\bibinfo
		{year} {2014})}\BibitemShut {NoStop}%
	\bibitem [{\citenamefont {Pan}\ \emph {et~al.}(2022)\citenamefont {Pan},
		\citenamefont {Yuan}, \citenamefont {Zhou}, \citenamefont {Zhang},
		\citenamefont {Li}, \citenamefont {Liu}, \citenamefont {Jiang}, \citenamefont
		{Catelani}, \citenamefont {Hu},\ and\ \citenamefont
		{Yan}}]{pan_2022_engineering}%
	\BibitemOpen
	\bibfield  {author} {\bibinfo {author} {\bibfnamefont {X.}~\bibnamefont
			{Pan}}, \bibinfo {author} {\bibfnamefont {H.}~\bibnamefont {Yuan}}, \bibinfo
		{author} {\bibfnamefont {Y.}~\bibnamefont {Zhou}}, \bibinfo {author}
		{\bibfnamefont {L.}~\bibnamefont {Zhang}}, \bibinfo {author} {\bibfnamefont
			{J.}~\bibnamefont {Li}}, \bibinfo {author} {\bibfnamefont {S.}~\bibnamefont
			{Liu}}, \bibinfo {author} {\bibfnamefont {Z.~H.}\ \bibnamefont {Jiang}},
		\bibinfo {author} {\bibfnamefont {G.}~\bibnamefont {Catelani}}, \bibinfo
		{author} {\bibfnamefont {L.}~\bibnamefont {Hu}},\ and\ \bibinfo {author}
		{\bibfnamefont {F.}~\bibnamefont {Yan}},\ }\bibfield  {title} {\bibinfo
		{title} {Engineering superconducting qubits to reduce quasiparticles and
			charge noise},\ }\href@noop {} {\bibfield  {journal} {\bibinfo  {journal}
			{arXiv:2202.01435}\ } (\bibinfo {year} {2022})}\BibitemShut {NoStop}%
	\bibitem [{\citenamefont {Karatsu}\ \emph {et~al.}(2019)\citenamefont
		{Karatsu}, \citenamefont {Endo}, \citenamefont {Bueno}, \citenamefont
		{De~Visser}, \citenamefont {Barends}, \citenamefont {Thoen}, \citenamefont
		{Murugesan}, \citenamefont {Tomita},\ and\ \citenamefont
		{Baselmans}}]{karatsu_2019_mitigation}%
	\BibitemOpen
	\bibfield  {author} {\bibinfo {author} {\bibfnamefont {K.}~\bibnamefont
			{Karatsu}}, \bibinfo {author} {\bibfnamefont {A.}~\bibnamefont {Endo}},
		\bibinfo {author} {\bibfnamefont {J.}~\bibnamefont {Bueno}}, \bibinfo
		{author} {\bibfnamefont {P.}~\bibnamefont {De~Visser}}, \bibinfo {author}
		{\bibfnamefont {R.}~\bibnamefont {Barends}}, \bibinfo {author} {\bibfnamefont
			{D.}~\bibnamefont {Thoen}}, \bibinfo {author} {\bibfnamefont
			{V.}~\bibnamefont {Murugesan}}, \bibinfo {author} {\bibfnamefont
			{N.}~\bibnamefont {Tomita}},\ and\ \bibinfo {author} {\bibfnamefont
			{J.}~\bibnamefont {Baselmans}},\ }\bibfield  {title} {\bibinfo {title}
		{Mitigation of cosmic ray effect on microwave kinetic inductance detector
			arrays},\ }\href@noop {} {\bibfield  {journal} {\bibinfo  {journal} {App.
				Phys. Lett.}\ }\textbf {\bibinfo {volume} {114}},\ \bibinfo {pages} {032601}
		(\bibinfo {year} {2019})}\BibitemShut {NoStop}%
	\bibitem [{\citenamefont {Riwar}\ and\ \citenamefont
		{Catelani}(2019)}]{riwar_2019_efficient}%
	\BibitemOpen
	\bibfield  {author} {\bibinfo {author} {\bibfnamefont {R.-P.}\ \bibnamefont
			{Riwar}}\ and\ \bibinfo {author} {\bibfnamefont {G.}~\bibnamefont
			{Catelani}},\ }\bibfield  {title} {\bibinfo {title} {Efficient quasiparticle
			traps with low dissipation through gap engineering},\ }\href@noop {}
	{\bibfield  {journal} {\bibinfo  {journal} {Phys. Rev. B}\ }\textbf {\bibinfo
			{volume} {100}},\ \bibinfo {pages} {144514} (\bibinfo {year}
		{2019})}\BibitemShut {NoStop}%
	\bibitem [{\citenamefont {Zhao}\ \emph {et~al.}(2022)\citenamefont {Zhao},
		\citenamefont {Ma}, \citenamefont {Jin},\ and\ \citenamefont
		{Yu}}]{zhao_2022_combating}%
	\BibitemOpen
	\bibfield  {author} {\bibinfo {author} {\bibfnamefont {P.}~\bibnamefont
			{Zhao}}, \bibinfo {author} {\bibfnamefont {T.}~\bibnamefont {Ma}}, \bibinfo
		{author} {\bibfnamefont {Y.}~\bibnamefont {Jin}},\ and\ \bibinfo {author}
		{\bibfnamefont {H.}~\bibnamefont {Yu}},\ }\bibfield  {title} {\bibinfo
		{title} {Combating fluctuations in relaxation times of fixed-frequency
			transmon qubits with microwave-dressed states},\ }\href@noop {} {\bibfield
		{journal} {\bibinfo  {journal} {Phys. Rev. A}\ }\textbf {\bibinfo {volume}
			{105}},\ \bibinfo {pages} {062605} (\bibinfo {year} {2022})}\BibitemShut
	{NoStop}%
	\bibitem [{\citenamefont {Paik}\ \emph {et~al.}(2011)\citenamefont {Paik},
		\citenamefont {Schuster}, \citenamefont {Bishop}, \citenamefont {Kirchmair},
		\citenamefont {Catelani}, \citenamefont {Sears}, \citenamefont {Johnson},
		\citenamefont {Reagor}, \citenamefont {Frunzio}, \citenamefont {Glazman}
		\emph {et~al.}}]{paik_2011_observation}%
	\BibitemOpen
	\bibfield  {author} {\bibinfo {author} {\bibfnamefont {H.}~\bibnamefont
			{Paik}}, \bibinfo {author} {\bibfnamefont {D.~I.}\ \bibnamefont {Schuster}},
		\bibinfo {author} {\bibfnamefont {L.~S.}\ \bibnamefont {Bishop}}, \bibinfo
		{author} {\bibfnamefont {G.}~\bibnamefont {Kirchmair}}, \bibinfo {author}
		{\bibfnamefont {G.}~\bibnamefont {Catelani}}, \bibinfo {author}
		{\bibfnamefont {A.~P.}\ \bibnamefont {Sears}}, \bibinfo {author}
		{\bibfnamefont {B.}~\bibnamefont {Johnson}}, \bibinfo {author} {\bibfnamefont
			{M.}~\bibnamefont {Reagor}}, \bibinfo {author} {\bibfnamefont
			{L.}~\bibnamefont {Frunzio}}, \bibinfo {author} {\bibfnamefont {L.~I.}\
			\bibnamefont {Glazman}}, \emph {et~al.},\ }\bibfield  {title} {\bibinfo
		{title} {Observation of high coherence in josephson junction qubits measured
			in a three-dimensional circuit qed architecture},\ }\href@noop {} {\bibfield
		{journal} {\bibinfo  {journal} {Phys. Rev. Lett.}\ }\textbf {\bibinfo
			{volume} {107}},\ \bibinfo {pages} {240501} (\bibinfo {year}
		{2011})}\BibitemShut {NoStop}%
	\bibitem [{\citenamefont {Bilmes}\ \emph {et~al.}(2020)\citenamefont {Bilmes},
		\citenamefont {Megrant}, \citenamefont {Klimov}, \citenamefont {Weiss},
		\citenamefont {Martinis}, \citenamefont {Ustinov},\ and\ \citenamefont
		{Lisenfeld}}]{bilmes_2020_resolving}%
	\BibitemOpen
	\bibfield  {author} {\bibinfo {author} {\bibfnamefont {A.}~\bibnamefont
			{Bilmes}}, \bibinfo {author} {\bibfnamefont {A.}~\bibnamefont {Megrant}},
		\bibinfo {author} {\bibfnamefont {P.}~\bibnamefont {Klimov}}, \bibinfo
		{author} {\bibfnamefont {G.}~\bibnamefont {Weiss}}, \bibinfo {author}
		{\bibfnamefont {J.~M.}\ \bibnamefont {Martinis}}, \bibinfo {author}
		{\bibfnamefont {A.~V.}\ \bibnamefont {Ustinov}},\ and\ \bibinfo {author}
		{\bibfnamefont {J.}~\bibnamefont {Lisenfeld}},\ }\bibfield  {title} {\bibinfo
		{title} {Resolving the positions of defects in superconducting quantum
			bits},\ }\href@noop {} {\bibfield  {journal} {\bibinfo  {journal} {Sci.
				Rep.}\ }\textbf {\bibinfo {volume} {10}},\ \bibinfo {pages} {1} (\bibinfo
		{year} {2020})}\BibitemShut {NoStop}%
	\bibitem [{\citenamefont {Nazaretski}\ \emph {et~al.}(2004)\citenamefont
		{Nazaretski}, \citenamefont {Merithew}, \citenamefont {Kostroun},
		\citenamefont {Zehnder}, \citenamefont {Pohl},\ and\ \citenamefont
		{Parpia}}]{nazaretski_2004_effect}%
	\BibitemOpen
	\bibfield  {author} {\bibinfo {author} {\bibfnamefont {E.}~\bibnamefont
			{Nazaretski}}, \bibinfo {author} {\bibfnamefont {R.}~\bibnamefont
			{Merithew}}, \bibinfo {author} {\bibfnamefont {V.}~\bibnamefont {Kostroun}},
		\bibinfo {author} {\bibfnamefont {A.}~\bibnamefont {Zehnder}}, \bibinfo
		{author} {\bibfnamefont {R.}~\bibnamefont {Pohl}},\ and\ \bibinfo {author}
		{\bibfnamefont {J.}~\bibnamefont {Parpia}},\ }\bibfield  {title} {\bibinfo
		{title} {Effect of low-level radiation on the low temperature acoustic
			behavior of {a-SiO$_2$}},\ }\href@noop {} {\bibfield  {journal} {\bibinfo
			{journal} {Phys. Rev. Lett.}\ }\textbf {\bibinfo {volume} {92}},\ \bibinfo
		{pages} {245502} (\bibinfo {year} {2004})}\BibitemShut {NoStop}%
	\bibitem [{\citenamefont {Anthony-Petersen}\ \emph {et~al.}(2022)\citenamefont
		{Anthony-Petersen}, \citenamefont {Biekert}, \citenamefont {Bunker},
		\citenamefont {Chang}, \citenamefont {Chang}, \citenamefont {Chaplinsky},
		\citenamefont {Fascione}, \citenamefont {Fink}, \citenamefont
		{Garcia-Sciveres}, \citenamefont {Germond} \emph
		{et~al.}}]{anthony2022stress}%
	\BibitemOpen
	\bibfield  {author} {\bibinfo {author} {\bibfnamefont {R.}~\bibnamefont
			{Anthony-Petersen}}, \bibinfo {author} {\bibfnamefont {A.}~\bibnamefont
			{Biekert}}, \bibinfo {author} {\bibfnamefont {R.}~\bibnamefont {Bunker}},
		\bibinfo {author} {\bibfnamefont {C.~L.}\ \bibnamefont {Chang}}, \bibinfo
		{author} {\bibfnamefont {Y.-Y.}\ \bibnamefont {Chang}}, \bibinfo {author}
		{\bibfnamefont {L.}~\bibnamefont {Chaplinsky}}, \bibinfo {author}
		{\bibfnamefont {E.}~\bibnamefont {Fascione}}, \bibinfo {author}
		{\bibfnamefont {C.~W.}\ \bibnamefont {Fink}}, \bibinfo {author}
		{\bibfnamefont {M.}~\bibnamefont {Garcia-Sciveres}}, \bibinfo {author}
		{\bibfnamefont {R.}~\bibnamefont {Germond}}, \emph {et~al.},\ }\bibfield
	{title} {\bibinfo {title} {A stress induced source of phonon bursts and
			quasiparticle poisoning},\ }\href@noop {} {\bibfield  {journal} {\bibinfo
			{journal} {arXiv:2208.02790}\ } (\bibinfo {year} {2022})}\BibitemShut
	{NoStop}%
	\bibitem [{\citenamefont {Gusenkova}\ \emph {et~al.}(2022)\citenamefont
		{Gusenkova}, \citenamefont {Valenti}, \citenamefont {Spiecker}, \citenamefont
		{G{\"u}nzler}, \citenamefont {Paluch}, \citenamefont {Rieger}, \citenamefont
		{Piora{\c{s}}-{\c{T}}imbolma{\c{s}}}, \citenamefont {Z{\^a}rbo},
		\citenamefont {Casali}, \citenamefont {Colantoni} \emph
		{et~al.}}]{gusenkova2022operating}%
	\BibitemOpen
	\bibfield  {author} {\bibinfo {author} {\bibfnamefont {D.}~\bibnamefont
			{Gusenkova}}, \bibinfo {author} {\bibfnamefont {F.}~\bibnamefont {Valenti}},
		\bibinfo {author} {\bibfnamefont {M.}~\bibnamefont {Spiecker}}, \bibinfo
		{author} {\bibfnamefont {S.}~\bibnamefont {G{\"u}nzler}}, \bibinfo {author}
		{\bibfnamefont {P.}~\bibnamefont {Paluch}}, \bibinfo {author} {\bibfnamefont
			{D.}~\bibnamefont {Rieger}}, \bibinfo {author} {\bibfnamefont {L.-M.}\
			\bibnamefont {Piora{\c{s}}-{\c{T}}imbolma{\c{s}}}}, \bibinfo {author}
		{\bibfnamefont {L.~P.}\ \bibnamefont {Z{\^a}rbo}}, \bibinfo {author}
		{\bibfnamefont {N.}~\bibnamefont {Casali}}, \bibinfo {author} {\bibfnamefont
			{I.}~\bibnamefont {Colantoni}}, \emph {et~al.},\ }\bibfield  {title}
	{\bibinfo {title} {Operating in a deep underground facility improves the
			locking of gradiometric fluxonium qubits at the sweet spots},\ }\href@noop {}
	{\bibfield  {journal} {\bibinfo  {journal} {App. Phys. Lett.}\ }\textbf
		{\bibinfo {volume} {120}},\ \bibinfo {pages} {054001} (\bibinfo {year}
		{2022})}\BibitemShut {NoStop}%
	\bibitem [{\citenamefont {Faoro}\ and\ \citenamefont
		{Ioffe}(2006)}]{faoro_2006_quantum}%
	\BibitemOpen
	\bibfield  {author} {\bibinfo {author} {\bibfnamefont {L.}~\bibnamefont
			{Faoro}}\ and\ \bibinfo {author} {\bibfnamefont {L.~B.}\ \bibnamefont
			{Ioffe}},\ }\bibfield  {title} {\bibinfo {title} {Quantum two level systems
			and kondo-like traps as possible sources of decoherence in superconducting
			qubits},\ }\href@noop {} {\bibfield  {journal} {\bibinfo  {journal} {Phys.
				Rev. Lett.}\ }\textbf {\bibinfo {volume} {96}},\ \bibinfo {pages} {047001}
		(\bibinfo {year} {2006})}\BibitemShut {NoStop}%
	\bibitem [{\citenamefont {Agarwal}\ \emph {et~al.}(2013)\citenamefont
		{Agarwal}, \citenamefont {Martin}, \citenamefont {Lukin},\ and\ \citenamefont
		{Demler}}]{agarwal_2013_polaronic}%
	\BibitemOpen
	\bibfield  {author} {\bibinfo {author} {\bibfnamefont {K.}~\bibnamefont
			{Agarwal}}, \bibinfo {author} {\bibfnamefont {I.}~\bibnamefont {Martin}},
		\bibinfo {author} {\bibfnamefont {M.~D.}\ \bibnamefont {Lukin}},\ and\
		\bibinfo {author} {\bibfnamefont {E.}~\bibnamefont {Demler}},\ }\bibfield
	{title} {\bibinfo {title} {Polaronic model of two-level systems in amorphous
			solids},\ }\href@noop {} {\bibfield  {journal} {\bibinfo  {journal} {Phys.
				Rev. B}\ }\textbf {\bibinfo {volume} {87}},\ \bibinfo {pages} {144201}
		(\bibinfo {year} {2013})}\BibitemShut {NoStop}%
	\bibitem [{\citenamefont {Lutchyn}\ \emph {et~al.}(2008)\citenamefont
		{Lutchyn}, \citenamefont {Cywi{\'n}ski}, \citenamefont {Nave},\ and\
		\citenamefont {Sarma}}]{lutchyn_2008_quantum}%
	\BibitemOpen
	\bibfield  {author} {\bibinfo {author} {\bibfnamefont {R.~M.}\ \bibnamefont
			{Lutchyn}}, \bibinfo {author} {\bibfnamefont {{\L}.}~\bibnamefont
			{Cywi{\'n}ski}}, \bibinfo {author} {\bibfnamefont {C.~P.}\ \bibnamefont
			{Nave}},\ and\ \bibinfo {author} {\bibfnamefont {S.~D.}\ \bibnamefont
			{Sarma}},\ }\bibfield  {title} {\bibinfo {title} {Quantum decoherence of a
			charge qubit in a spin-fermion model},\ }\href@noop {} {\bibfield  {journal}
		{\bibinfo  {journal} {Phys. Rev. B}\ }\textbf {\bibinfo {volume} {78}},\
		\bibinfo {pages} {024508} (\bibinfo {year} {2008})}\BibitemShut {NoStop}%
	\bibitem [{\citenamefont {Choi}\ \emph {et~al.}(2009)\citenamefont {Choi},
		\citenamefont {Lee}, \citenamefont {Louie},\ and\ \citenamefont
		{Clarke}}]{choi_2009_localization}%
	\BibitemOpen
	\bibfield  {author} {\bibinfo {author} {\bibfnamefont {S.}~\bibnamefont
			{Choi}}, \bibinfo {author} {\bibfnamefont {D.-H.}\ \bibnamefont {Lee}},
		\bibinfo {author} {\bibfnamefont {S.~G.}\ \bibnamefont {Louie}},\ and\
		\bibinfo {author} {\bibfnamefont {J.}~\bibnamefont {Clarke}},\ }\bibfield
	{title} {\bibinfo {title} {Localization of metal-induced gap states at the
			metal-insulator interface: origin of flux noise in squids and superconducting
			qubits},\ }\href@noop {} {\bibfield  {journal} {\bibinfo  {journal} {Phys.
				Rev. Lett.}\ }\textbf {\bibinfo {volume} {103}},\ \bibinfo {pages} {197001}
		(\bibinfo {year} {2009})}\BibitemShut {NoStop}%
	\bibitem [{\citenamefont {Bespalov}\ \emph {et~al.}(2016)\citenamefont
		{Bespalov}, \citenamefont {Houzet}, \citenamefont {Meyer},\ and\
		\citenamefont {Nazarov}}]{bespalov_2016_theoretical}%
	\BibitemOpen
	\bibfield  {author} {\bibinfo {author} {\bibfnamefont {A.}~\bibnamefont
			{Bespalov}}, \bibinfo {author} {\bibfnamefont {M.}~\bibnamefont {Houzet}},
		\bibinfo {author} {\bibfnamefont {J.~S.}\ \bibnamefont {Meyer}},\ and\
		\bibinfo {author} {\bibfnamefont {Y.~V.}\ \bibnamefont {Nazarov}},\
	}\bibfield  {title} {\bibinfo {title} {Theoretical model to explain excess of
			quasiparticles in superconductors},\ }\href@noop {} {\bibfield  {journal}
		{\bibinfo  {journal} {Phys. Rev. Lett.}\ }\textbf {\bibinfo {volume} {117}},\
		\bibinfo {pages} {117002} (\bibinfo {year} {2016})}\BibitemShut {NoStop}%
	\bibitem [{\citenamefont {de~Graaf}\ \emph {et~al.}(2020)\citenamefont
		{de~Graaf}, \citenamefont {Faoro}, \citenamefont {Ioffe}, \citenamefont
		{Mahashabde}, \citenamefont {Burnett}, \citenamefont {Lindstr{\"o}m},
		\citenamefont {Kubatkin}, \citenamefont {Danilov},\ and\ \citenamefont
		{Tzalenchuk}}]{deGraaf_2020_qpTLS}%
	\BibitemOpen
	\bibfield  {author} {\bibinfo {author} {\bibfnamefont {S.}~\bibnamefont
			{de~Graaf}}, \bibinfo {author} {\bibfnamefont {L.}~\bibnamefont {Faoro}},
		\bibinfo {author} {\bibfnamefont {L.}~\bibnamefont {Ioffe}}, \bibinfo
		{author} {\bibfnamefont {S.}~\bibnamefont {Mahashabde}}, \bibinfo {author}
		{\bibfnamefont {J.}~\bibnamefont {Burnett}}, \bibinfo {author} {\bibfnamefont
			{T.}~\bibnamefont {Lindstr{\"o}m}}, \bibinfo {author} {\bibfnamefont
			{S.}~\bibnamefont {Kubatkin}}, \bibinfo {author} {\bibfnamefont
			{A.}~\bibnamefont {Danilov}},\ and\ \bibinfo {author} {\bibfnamefont {A.~Y.}\
			\bibnamefont {Tzalenchuk}},\ }\bibfield  {title} {\bibinfo {title} {Two-level
			systems in superconducting quantum devices due to trapped quasiparticles},\
	}\href@noop {} {\bibfield  {journal} {\bibinfo  {journal} {Sci. Adv.}\
		}\textbf {\bibinfo {volume} {6}},\ \bibinfo {pages} {eabc5055} (\bibinfo
		{year} {2020})}\BibitemShut {NoStop}%
\end{thebibliography}



\appendix


\section{Automatic detection of offset-charge jumps}
\label{app_algo}

\begin{figure}[h]
\includegraphics[width = \columnwidth]{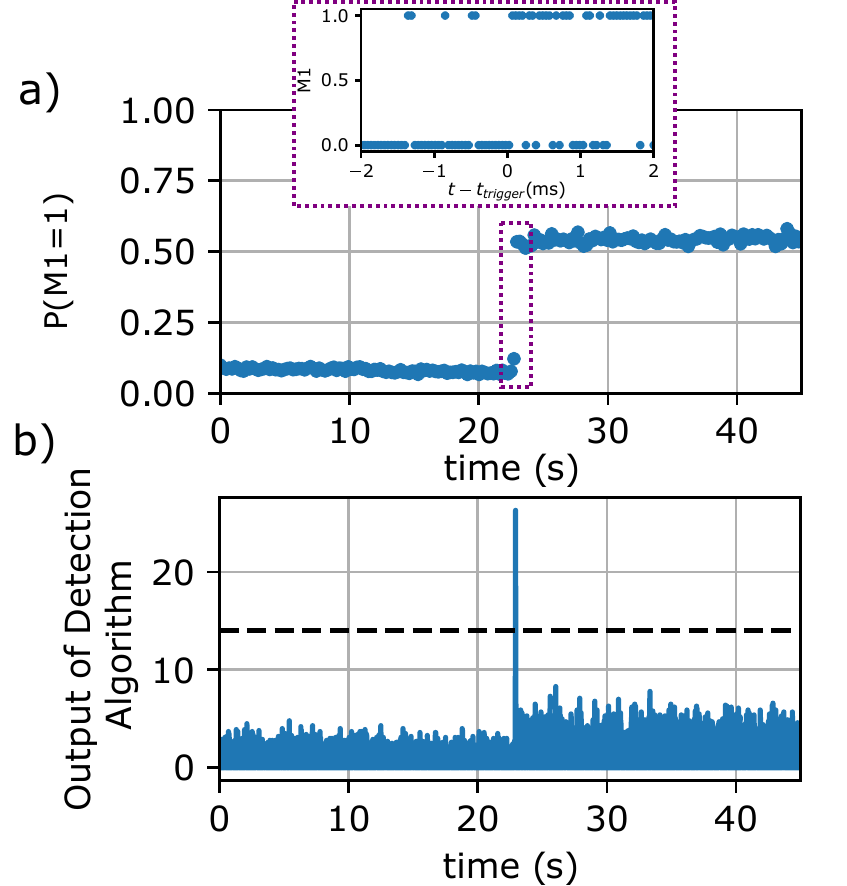}
\caption{Jump detection algorithm. a) Output of Ramsey-based jump detector, reproduced from Fig. \ref{fig_ng_jumps}(d).  In this panel, the 1 million shots of the experiment have been binned into 200 bins to calculated $P(M1\!=\!1)$. (Inset) Raw output of M1 for 2~ms before and after the jump.  b) Output of the jump detection algorithm after normalization by the median level, showing a peak (which sets $t_{trigger}$) corresponding to the jump in (a).  The dashed line is our default detection threshold of 14.}
\label{fig_supp_algo}
\end{figure}

Our experiments involved monitoring many qubits for hours at a time, and the large amounts of generated data required efficient, robust detection algorithms.  The output of the Ramsey jump detector from Fig. \ref{fig_ng_jumps}(d) is reproduced in Fig. \ref{fig_supp_algo}(a).  For plotting $P(M1\!=\!1)$, the data have been binned into 200 bins. However, binning discards precise timing information about the jump. The inset of Fig. \ref{fig_supp_algo}(a) shows the outcomes of the individual measurements immediately before and after the jump. By eye we can see that we should be able to detect the time of this jump to within a few time steps. To detect jumps and extract the time of each jump we used a matched filter with a step function with a duration of 50~ms as the template 
\[ \textrm{Template}(t) = \begin{cases} 
      -1, & -25~\textrm{ms} < t < 0 \\
      0, & t=0 \\
      1, & 0< t < 25~\textrm{ms}
   \end{cases}
\]
The cross-correlation (\textit{scipy.signal.correlate}) of the raw output of the jump detector and the template is the result of sliding the template along the output of the jump detector.  Absent a jump in the data, the positive and negative halves of the step function null the correlation, while a jump causes a spike in the correlation. The problem of step-detection thus reduces to peak-detection in the jump signal, defined as the absolute value of this cross-correlation, which we performed via thresholding. We chose this matched filter approach for simplicity; other techniques for offline changepoint detection can likely improve sensitivity in future studies (e.g. information might be gleaned not only from the step in mean value when $n_{g0}$ jumps but also from the step in variance). As the analyses of the 500 experiment runs are independent, time can be saved with parallel processing.

The choice of threshold contains two subtleties: how to normalize signals for cross-qubit comparisons, and then how to set the overall threshold level. Instantaneous values of $n_{g0}$ and $T_{2,ef}^*$ will vary over time and across qubits, causing the sensitivity of each qubit to fluctuate over time.  We found it effective to normalize the jump signal trace from each qubit for each run by its median, such that a given threshold value roughly corresponds to a fixed confidence in (SNR of) the jump signal, rather than to a uniform magnitude of the jump signal.  Then, we chose an overall threshold of 14, which we empirically found provided good confidence in and time resolution of events while still accepting enough data for further statistical analysis (details in App. \ref{app_algo_thresh}). An example output of this algorithm with a large jump signal is shown in Fig. \ref{fig_supp_algo}(b).  A clear peak well above the threshold (black dashed line) indicates the single-qubit jump.  The time step of this peak set $t_{trigger}$. Double-counting of a noisy threshold crossing was avoided by setting a minimum time separation of 50~ms between detections on the same qubit, keeping the largest peak in such cases (\textit{scipy.signal.find\_peaks}).

Finally, the identified single-qubit jump were further clustered into multi-qubit jumps.  Each jump was sorted by $t_{trigger}$ into a list.  Multi-qubit jumps were identified when the time between $t_{trigger}$ on the list was less than 10~ms.  We caution that in principle this algorithm permits an edge case where a chain of $<10$~ms delays generates an arbitrarily long single multi-qubit jump, but in our data all multi-qubit jump durations were below 2~ms (median duration $<400~\mu$s) for our choice of threshold = 14. With that threshold, we detected 43 multi-qubit jumps, consisting of 108 single-qubit jumps.

\subsection{Spatial and temporal jump correlations}
\label{app_algo_multiq}

How can we be sure that the events we identify as multi-qubit jumps are true multi-qubit jumps and not independent single-qubit jumps, that happened by coincidence to jump within 10~ms of each other?  We support our interpretation of detected jumps as resulting from real charging events on the chip by looking for expected correlations in space and time.

\begin{figure}[h]
\includegraphics[width = 0.9\columnwidth]{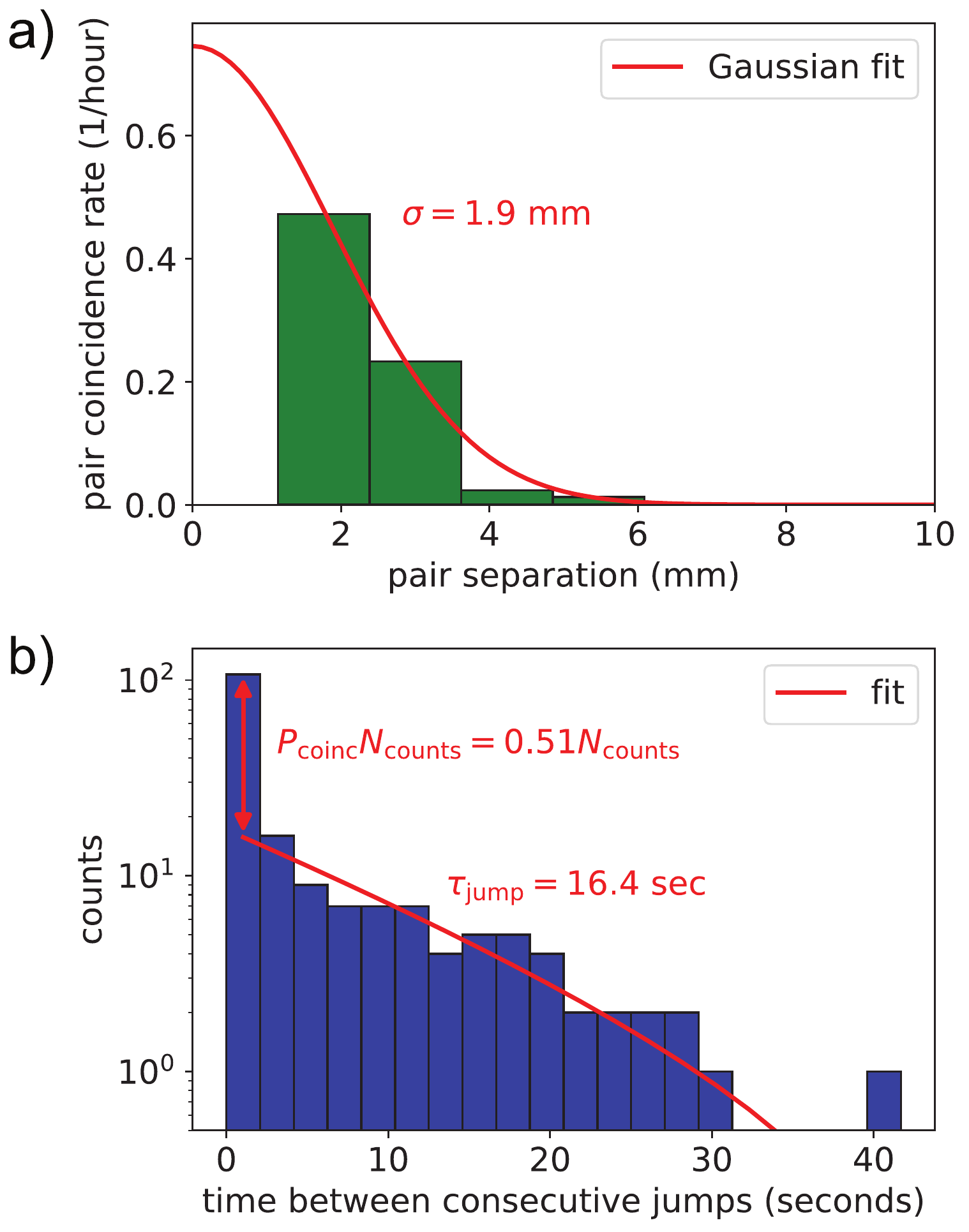}
\caption{Offset-charge jump correlations in space and time.  a) Histogram of the rate of simultaneous jumps for all pairs of physical qubits used for jump detection.  The first bin is empty because there is a minimal distance of about 1~mm between pairs of qubits.  In red is a Gaussian distribution, with $\sigma$~=~1.9~mm, showing that qubits that experience simultaneous jumps are located within a few millimeters of each other.  
b)  Histogram of the time between consecutive jumps (includes single-qubit jumps not associated with a multi-qubit jump).  The red line is a fit to a modified Poisson distribution, with a characteristic jump rate = 1/(16.4~s).  The fit includes a correction to account for the finite duration of the experiment, 44~s.  The Poisson distribution fails to account for the excess of events in the first bin, which accounts for 51~\% of total single-qubit jumps, indicating that the multi-qubit jumps cannot be explained as random coincidences of uncorrelated single-qubit jumps.}
\label{fig_coincidence}
\end{figure}

First, multi-qubit jumps with a common physical origin should preferentially involve qubits that are physically closer together.  If the multi-qubit jumps were random chance, distant qubits would be equally like to have simultaneous jumps as nearby qubits. For each pair of qubits used for jump detection, we calculated their separation and the rate at which the two qubits experienced a simultaneous jump (in units of multi-qubit jump events per hour of detector time).  In Fig. \ref{fig_coincidence}(a) we show a histogram of the coincidence rate as a function of the distance between qubits.  The distribution fits a Gaussian falloff with a $\sigma$ = 1.9~mm, indicating that the events are strongly localized.  Correlated offset-charge jumps have not been previously reported at this length scale \cite{wilen_correlated_2021, pan_2022_engineering},  though the characteristic length scale is expected to vary from device to device depending on the qubit design and chip layout.

Second, the time separation of independent random single-qubit jumps should be well described by a Poisson distribution, but multi-qubit jumps should instead exhibit more frequent coincidences where several qubits jump at once.  For each 44 second run of the jump detection experiment, we generated a single list of the times at which any individual qubit jumped, and calculated the time delays between consecutive single-qubit jumps within that run. During multi-qubit jumps, this time difference will be small, and can be zero. Figure \ref{fig_coincidence}(b) shows a histogram of these time differences with 2~s bins. If each jump were an independent single-qubit event, the Poisson-distributed delays between events would follow an exponential probability distribution with some characteristic timescale $t_\textrm{jump}$. However, the histogram shows a clear excess of events in the first bin, which cannot be explained by individual qubits independently experiencing jumps.

While these results are qualitatively consistent with expectations, further analysis is complicated by the possibility of multiple mechanisms contributing to jump detections.
In particular, single-qubit jumps can be explained by local processes not involving radiation \cite{stewart_2016_stability}.  For example, qubit 17 produced jumps in this data set at a higher rate than did the other qubits, likely due to a defect coupled to the ef-transition, so this outlier qubit was excluded from this rate analysis. Some fraction of the single-qubit jumps from the included qubits are likely also produced by causes other than radiation.
Nonetheless, we can grossly quantify the full distribution using the fit model
$$ \textrm{PDF}(\Delta) = N e^{-\Delta/\tau_\textrm{jump}} (1- \Delta/T)(1-P_\textrm{coinc}) + P_\textrm{coinc}\delta_\Delta$$
where PDF stands for probability density function, $\Delta$ is the time between events, $N = \frac{1}{\tau_\textrm{jump}} - \frac{1 - e^{-T/\tau_\textrm{jump} }}{T}$ is a normalization constant, and the factor $1- \frac{\Delta}{T} $ accounts for the finite duration $T = 44~$s  of the experiment. The free parameters are $\tau_\textrm{jump}$ and $P_\textrm{coinc}$, representing the excess probability that a single-qubit jumps occurs in coincidence with the preceding single-qubit jump (here, the excess probability mass in the first histogram bin). To fit the histogram, the PDF must be converted to a probability mass function by integrating over each bin, where we define $\delta_\Delta$ to integrate to 1 in the first bin and zero otherwise. The simplest interpretation of the fit result is that events producing jumps above threshold occurred with a timescale of 16.4 seconds, and many of those events caused more than one qubit to jump, increasing the number of jump counts in the first bin roughly tenfold from the expectation if qubits were jumping independently.

\subsection{Comparison of rate to literature}
\label{app_comp_rate}
If we assume that all of the jumps that we observe are due to radiation, then we can place a bound on the rate of the impact of radiation on the chip, and we can compare the this rate to other reports for the rate of radiation impacts from the literature, as shown in Table. \ref{tab:rate_table}.  The rate of impact from radiation will depend on many physical parameters, including the amount and type of local radioactivity, the amount of shielding, and the thickness of the substrate. Remarkably, dividing the average impact rates by the relevant chip sizes yields values that are all within roughly a factor of two.  This is striking agreement considering that these experiments used both different hardware technologies and detection techniques, which further supports the interpretation of our jumps as resulting from radiation. It also suggests that other differences in effects that we see, such as the effect on $T_1$, cannot be explained by a difference in the rate of events.

\begin{table*}[]
    \centering
    \begin{tabular}{| c| c |c |c |c |c | }
    \hline
     \textbf{Reference} & \textbf{Technology} & \textbf{Measurement} & \textbf{Time between} & \textbf{Size of Chip}  & \textbf{Normalized Rate} \\
     & & & \textbf{Impacts (s)} & \textbf{(mm$^2$)} & \textbf{($10^{-3}$/(s mm$^2$))} \\\hline
     This work & Transmons & offset-charge & 16 & 150 & 0.4 \\ \hline
     Wilen \cite{wilen_correlated_2021} & Transmons & offset-charge & 50 & 39 & 0.5 \\ \hline
     McEwen \cite{mcewen_resolving_2021} & Transmons & Correlated decays & 10& 100 & 1 \\ \hline
     Cardani \cite{cardani_reducing_2021} & grAl Resonators & Resonator frequency shift & 10 & 120 & 0.8 \\ \hline
     Gr\"{u}nhaupt \cite{grunhaupt_loss_2018} & grAl Resonators & Resonator frequency shift & 20 & 120 & 0.4 \\ \hline
    \end{tabular}
    \caption{Comparison of impact rates of radiation from the literature, showing that the rate of events normalized by the area of the chip that we observe is within a factor of two of other reports in the literature. }
    \label{tab:rate_table}
\end{table*}

\subsection{Alternative thresholds}
\label{app_algo_thresh}

\begin{figure*}[t!]
	\centering
	\includegraphics[width=\textwidth]{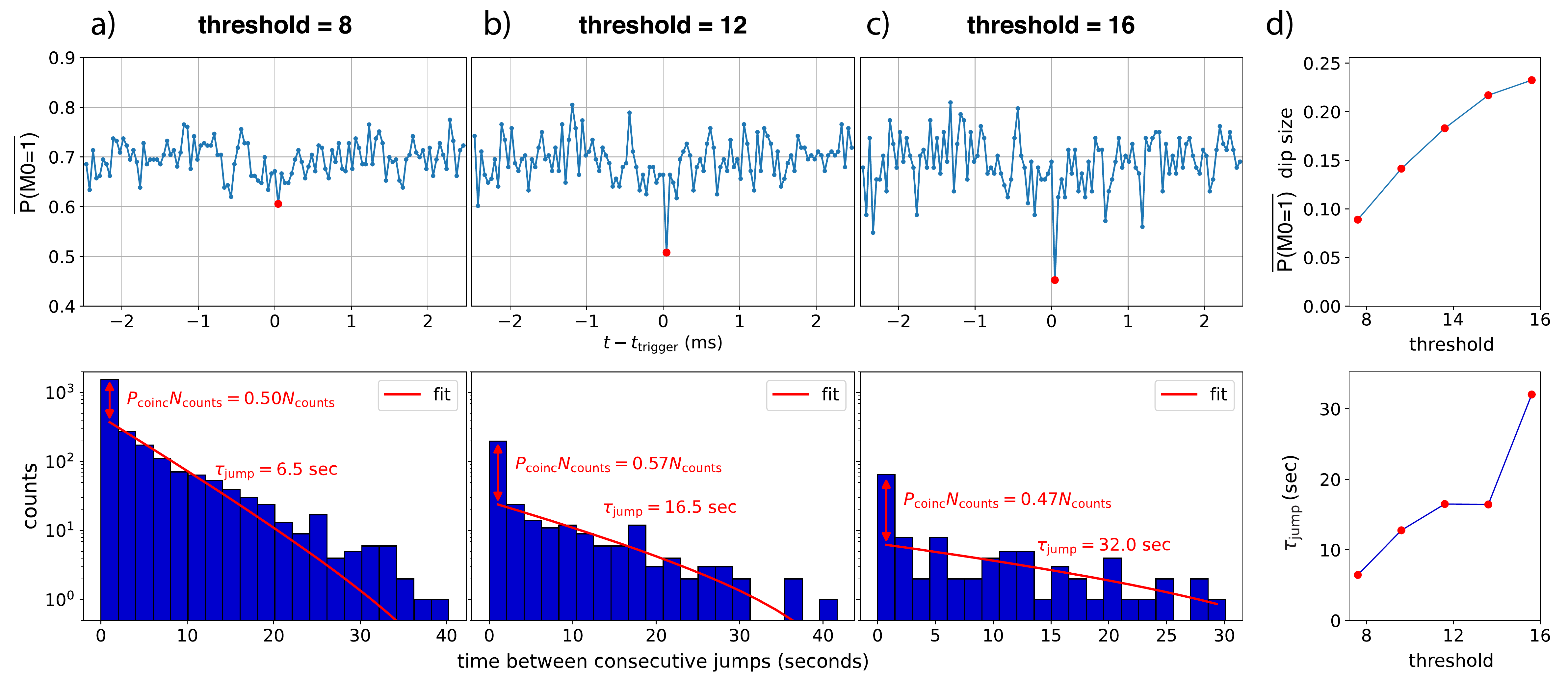}
    \caption{Comparisons of various thresholds for the jump detection. (a-c) The top panels show the equivalent of Fig. \ref{fig_qp_t1}(a) for increasing jump detection thresholds of 8, 12, and 16. (d) The size of the dip increases with threshold, perhaps due to tighter time resolution of the jump detection.  The bottom panels of (a-c) show the equivalent of Fig. \ref{fig_coincidence}(b).  Increasing the detection threshold decreases the characteristic rate at which events satisfying that threshold occur.}
    \label{fig_threshold_comparison}
\end{figure*}

We deliberately chose a high threshold of 14 for the jump detection algorithm to suppress false-positive detections, especially given that the noise properties of the device are likely to fluctuate appreciably on the multi-hour timescale of the experiment. However, we are free to reanalyze the same data with other thresholds to learn how various quantities systematically vary with threshold choice.

The top row of Fig. \ref{fig_threshold_comparison}(a-c) shows the equivalent of Fig. \ref{fig_qp_t1} for thresholds ranging from 8 to 16. Interestingly, the dip in $T_1$ becomes less prominent as the threshold is lowered, as shown in Fig. \ref{fig_threshold_comparison}(d). For the smallest threshold, Fig. \ref{fig_threshold_comparison}(a), the dip all but vanishes.  There are two possible explanations for this disappearance.  First, a higher threshold might select for events that generate more QPs, leading to a more prominent dip in $T_1$. Alternatively, keeping only jumps with a large, clear change in $P(MR\!=\!1)$ during the offset-charge jump detector experiment should reduce uncertainty in the timing of the jump, improving resolution of the dip.  To assess the timing accuracy of the jump detector, we fed numerically simulated results of the experiment into the jump detector algorithm (App. \ref{app_algo_timing}).

The bottom row of Fig. \ref{fig_threshold_comparison}(a-c) show the equivalent of Fig. \ref{fig_coincidence}(b) for the various thresholds. As expected when we raise the threshold we see the rate of events decline, from 6.5~s between impacts to 32~s between impacts. Deviation of the threshold-8 histogram from the modified Poisson fit may indicate an increasing significance of false-positive detections. At the other extreme, the threshold-16 histogram exhibits a scarcity of detections inhibiting statistical analysis.  In Table \ref{tab:rate_table} we use the rate $\tau_{\textrm{jump}}$ = 16.4~s, using the threshold-14 that we used in the main text.  As we vary the threshold, $\tau_{\textrm{jump}}$ can change by a factor of 2, but this does not significantly affect the comparison.

\subsection{Simulated timing accuracy of jump detector}
\label{app_algo_timing}

Figs. \ref{fig_qp_t1} \& \ref{fig_threshold_comparison} show a reduction in $T_1$ after an impact which consists of a single data point.  However, the Ramsey-based jump detector is probabilistic, so it is not obvious that the time of the jump can be determined to within a single time step even in principle.  To probe the feasible timing accuracy of the jump detector algorithm, we generated data from a model simulation of the experiment. In the simulation, we generated a list of impact times from a Poisson distribution of rate 1/(10~s), randomly selected coordinates on the chip for each impact, and calculated the magnitude of $\Delta n_{g0}$ for each impact based on the distance of each qubit to the impact site assuming a maximum $\Delta n_{g0}$~=~0.1 and a Gaussian falloff using $\sigma$ = 1.5~mm.  For each qubit, $n_{g0}$ was initialized to a uniformly random value, and slowly diffused with a variance of 0.1 electron/hour. Phase flips during Ramsey occurred per a coherence time of the $ef$ subspace $T_{2,ef}^*$ = 50~$\mu$s, and each qubit measurement had a symmetric error probability of 1.5~\%. Finally, to permit comparison with Fig. \ref{fig_qp_t1}, each simulated impact reduces $T_1$ of nearby qubits for a single timepoint, lowering $T_1$ to $1~\mu$s at the epicenter with the same Gaussian spatial falloff of $\sigma$ = 1.5~mm.

\begin{figure}[t!]
\centering
\includegraphics[width = 0.4\textwidth]{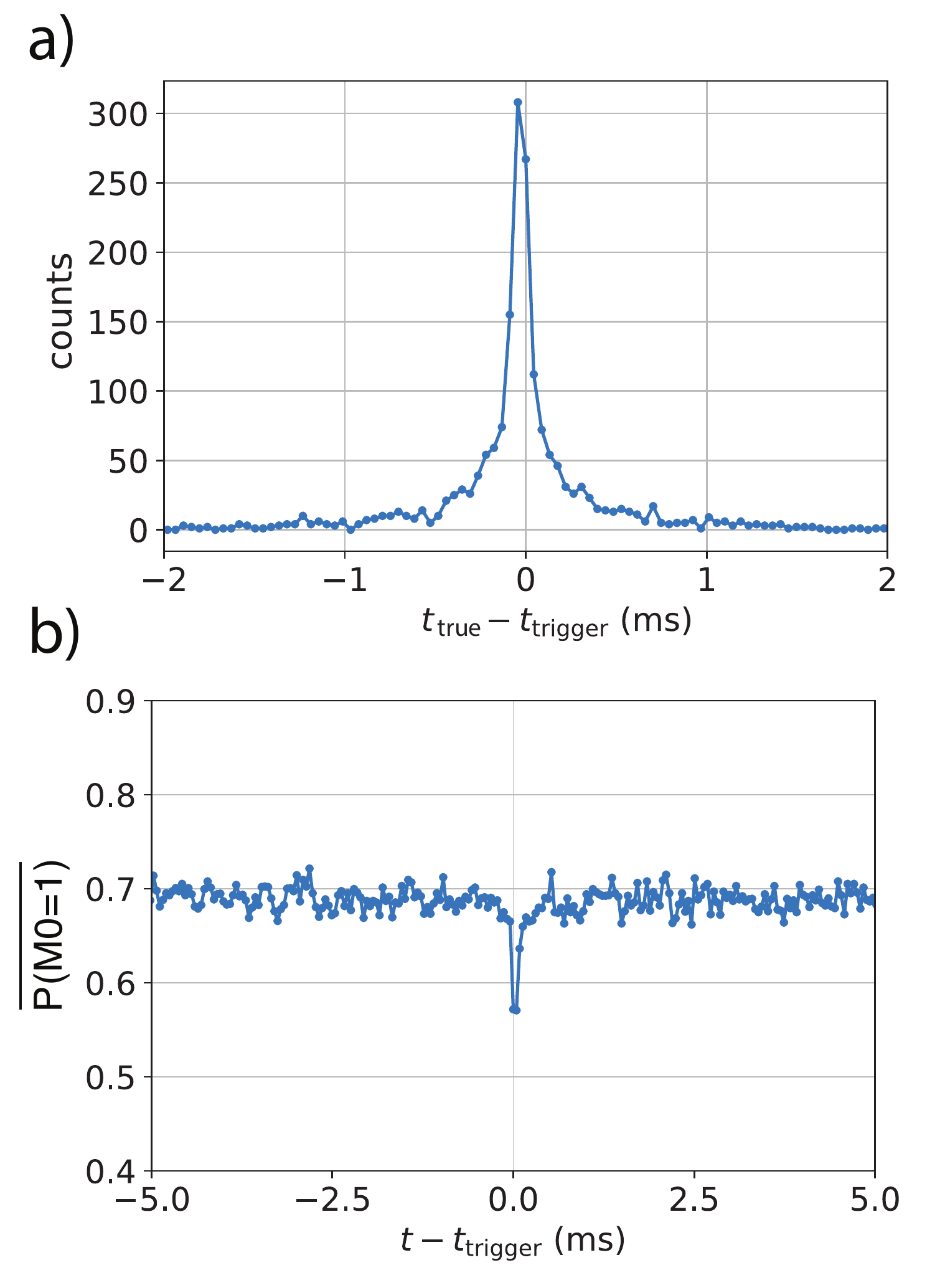}
\caption{jump detection of numerically simulated data, for threshold-14. (a) Shows histograms of the time differences between the times at which the analysis detected a jump event, and the true times of the corresponding simulated events.  Each bin corresponds to a single time step (44~$\mu$s).  (b) Shows the dip resulting from a transient decrease in $T_1$, for comparison with Fig. \ref{fig_qp_t1}. These results do not reproduce the single-point feature in Fig. \ref{fig_qp_t1}, but do suggest that few-point resolution is possible.}
\label{fig_ng_detector_spread}
\end{figure}

We ran the simulation 250 times, half as many as in the actual experiment, with the results shown in Fig. \ref{fig_ng_detector_spread}.  Because this is a simulation, we can compare the true time of the jump, $t_{true}$ to the time of the jump extracted from the jump detection algorithm, $t_{trigger}$ in  Fig. \ref{fig_ng_detector_spread}(a). Although the tails of the distribution are very long, far from either a Gaussian or a Lorentzian distribution, the extracted jump times are closely peaked around the true event time, such that very precisely resolving the time of the jump is possible. The resolution is better described by the median absolute deviation 132~$\mu$s (which is three time steps), than by the standard deviation 746~$\mu$s. Figure \ref{fig_ng_detector_spread}(b) shows the equivalent of Fig. \ref{fig_qp_t1}, for the simulated data.  The spread in $\overline{P(M0\!=\!1)}$ is reduced from the experiment, because the simulation does not account for all sources of variation, such as a spread in qubit $T_1$s.  Notably we do not observe single time step accuracy, as needed to explain Figs. \ref{fig_qp_t1} and \ref{fig_threshold_comparison}.  This may still arise from discrepancies between simulated and experimental error processes, or perhaps the noise in Fig. \ref{fig_qp_t1} could be obscuring the full width of the dip, in which case future improvements in qubit coherence may help reveal the full duration of the transient dynamics. 

\section{Automating detection of TLS scrambling}
\label{app_algo_scramble}
For the TLS scrambling experiment we ran the pulse sequence in Fig. \ref{fig_stark_TLS}(a), which used a variation on the jump detection sequence.    The fundamental difference is now the Ramsey jump detector experiment is run back to back 251 times in between sweeps of the TLS spectra.  Because the Stark shift measurements occupy three quarters of the experimental time, most of the radiation impact events will occur in between runs of the Ramsey jump detector, so the very precise temporal resolution of the jump detection algorithm from App. \ref{app_algo_timing} is irrelevant. Instead we averaged those 251 shots to calculate $P(MR\!=\!1)$ before running a slightly modified jump detection algorithm.  We used the same template as in App. \ref{app_algo_multiq}, except with a duration of the template of 200 sequences (5~s), to match the averaging time for the Pearson's $r$ calculation in the main text. By feeding the averaged $P(MR\!=\!1)$ into the jump detection algorithm rather than the raw output of the Ramsey sequence, we greatly increased the sensitivity to small jumps, at the cost of decreasing the temporal resolution of the extracted jump time to the time to run a single sequence (about 24~ms). We were able to use a smaller threshold of 8 to detect a jump, because the signal was less noisy.    For this reason we detect more multi-qubit jumps per hour than in App. \ref{app_algo_multiq}, as what would have been classified as single-qubit jumps were here classified as multi-qubit jumps.

\end{document}